\documentclass[12pt]{article}
\usepackage{ametsoc}
\pdfoutput=1
\newcommand{\myabstract}{}
\begin{document}
%
%%%%%%%%%%%%%%%%%%%%%%%%%%%%%%%%%%%%%%%%%%%%%%%%%%%%%%%%%%%%%%%%%%%%%
% TITLE
%
% Enter your TITLE here
%%%%%%%%%%%%%%%%%%%%%%%%%%%%%%%%%%%%%%%%%%%%%%%%%%%%%%%%%%%%%%%%%%%%%
\title{\textbf{\large{Baroclinic stationary waves in aquaplanet models}}}
%
% Author names, with corresponding author information. 
% [Update and move the \thanks{...} block as appropriate.]
%
\author{\textsc{Giuseppe Zappa}
				\thanks{\textit{Corresponding author address:} 
				Giuseppe Zappa, Centro Euro Mediterraneo per i Cambiamenti Climatici, 
				Viale Aldo Moro 44, 40127 Bologna, Italy
				\newline{E-mail: giuseppe.zappa@cmcc.it}}\\
\centerline{\textit{\footnotesize{Science and Management of Climate Change Program, Ca'Foscari University, Venice, Italy}}}\\
\centerline{\textit{\footnotesize{Centro Euro--Mediterraneo per i Cambiamenti Climatici, Bologna, Italy}}}\\
\and 
\centerline{\textsc{Valerio Lucarini}}\\% Add additional authors, different insitution
\centerline{\textit{\footnotesize{Department of Meteorology, University of Reading, Reading, UK}}}\\
\centerline{\textit{\footnotesize{Department of Mathematics, University of Reading, Reading, UK}}}\\
\and
\centerline{\textsc{Antonio Navarra}}\\% Add additional authors, different insitution
\centerline{\textit{\footnotesize{Centro Euro--Mediterraneo per i Cambiamenti Climatici, Bologna, Italy}}}\\
\centerline{\textit{\footnotesize{Istituto Nazionale di Geofisica e Vulcanologia, Bologna, Italy}}}\\
}
%
% The following block of code will handle the formatting of the title page depnding on whether
% we are formatting a double column (dc) author draft or a single column paper for submission.
% AUTHORS SHOULD SKIP OVER THIS... There is nothing to do in this section of code.
\ifthenelse{\boolean{dc}}
{
\twocolumn[
\begin{@twocolumnfalse}
\amstitle

% Start Abstract (Enter your Abstract above.  Do not enter any text here)
\begin{center}
\begin{minipage}{13.0cm}
\begin{abstract}
	\myabstract
	\newline
	\begin{center}
		\rule{38mm}{0.2mm}
	\end{center}
\end{abstract}
\end{minipage}
\end{center}
\end{@twocolumnfalse}
]
}
{
\amstitle
\begin{abstract}
\myabstract

An aquaplanet model is used to study the nature of the highly persistent low frequency waves that have been observed in models forced by zonally symmetric boundary conditions. 

Using the Hayashi spectral analysis of the extratropical waves, we find that a quasi--stationary wave five belongs to a  wave packet obeying a well defined dispersion relation with eastward group velocity. The components of the dispersion relation with $k \geqslant 5$ baroclinically convert  eddy available potential energy into eddy kinetic energy, while those with $k<5$ are baroclinically neutral. In agreement with the Green's model of baroclinic instability, the wave five is weakly  unstable, and the inverse energy cascade, which had been previously proposed as a main forcing for this type of waves, only acts as a positive feedback on its predominantly baroclinic energetics. The quasi--stationary wave is reinforced by a phase lock to an analogous pattern in the tropical convection, which provides further amplification to the wave. We also find that the Pedlosky bounds on the phase speed of unstable waves provide guidance in  explaining  the latitudinal structure of the energy conversion, which is shown to be more enhanced where the zonal westerly surface wind is weaker. The wave's energy is then trapped in the wave guide created by the upper tropospheric jet stream. In agreement with Green's theory, as the equator to pole SST difference is reduced the stationary marginally stable component shifts toward higher wavenumbers, while the wave five becomes neutral and westward propagating. 

Some properties of the aquaplanet quasi--stationary waves are found in interesting agreement with a low frequency wave observed by \cite{Salby:1982} in the southern hemisphere DJF, so that this perspective on low frequency variability might be, apart from its value in terms of basic geophysical fluid dynamics, of specific interest for studying the Earth's atmosphere.

\end{abstract}
\newpage
}
%%%%%%%%%%%%%%%%%%%%%%%%%%%%%%%%%%%%%%%%%%%%%%%%%%%%%%%%%%%%%%%%%%%%%
% MAIN BODY OF PAPER
%%%%%%%%%%%%%%%%%%%%%%%%%%%%%%%%%%%%%%%%%%%%%%%%%%%%%%%%%%%%%%%%%%%%%
%

\section{Introduction}

Understanding the mechanisms that generate and maintain the atmospheric extratropical low frequency variability (LFV) is of primary importance for studying the basic properties of the climate system \citep{Benzi:1986, Vautard:1988, Benzi:1989, Branstator:1992,Haines:1994,Feldstein:1998,Itoh:1999, Dandrea:2002, Ruti:2006}, for evaluating the atmospheric predictability and extended weather range forecasts \citep{Palmer:1999}, and for detecting climate change signals as perturbations in the frequency of occurrence of the weather regimes \citep{Corti:1999}. 

The development of a self consistent theory of the LFV is complicated by the number of different, but interacting, processes that underlie its dynamics, so that simplified models have been introduced in order to cope with only a limited number of mechanisms \citep{Swanson:2002}. Many of these mechanisms, such as the baroclinic--orographic resonance via the form drag \citep{Benzi:1986, Ruti:2006}, the barotropic instability of the stationary waves \citep{Simmons:1983}, and the Rossby wave radiation from anomalous tropical convection \citep{Hoskins:1981}, require the presence of zonal asymmetries in the forcing of the mean state. LFV is, however, observed also in aquaplanets, which consist of GCMs set in a zonally symmetric ocean covered world. 

In typical extratropical settings, aquaplanets have been observed to feature a strong and persistent quasi--stationary zonal wave number five \citep{Watanabe:2005}, which has recently been found (Williamson 2009, personal communication) in most of the models participating in the Aquaplanet Intercomparison Project \citep{Neale:2001}. The matter is of interest because wave five quasi--stationary circumglobal patterns are also  observed along the jet stream in summer SH \citep{Salby:1982, Kidson:1999}, and in the NH during both the boreal winter  \citep{Branstator:2002, Chen:2002} and the summer \citep{Ding:2005,Blackburn:2008} seasons. While the strong wave guiding effect provided by the jet is known to be responsible for the zonal orientation of these teleconnection patterns, their dynamics is still not well understood, and the leaky normal mode theory proposed for the SH \citep{Salby:1982,Lin:1989}  has no clear counterpart in the NH. The attractive possibility to use aquaplanet models  as interpretative tools of these low frequency modes calls for a deeper understanding of their properties. 

A low frequency wavenumber five had been previously observed  by \cite{Hendon:1985} (hereafter HH) in a two level dry primitive equation model with zonally symmetric boundary conditions, and forced by newtonian relaxation. The constant presence of a dominant low frequency wave five in a variety of models and for different boundary conditions suggests that there must be a fundamental atmospheric process leading to its formation.  HH proposed a QG inverse turbulent  energy cascade \citep{Charney:1971, Rhines:1975, Basdevant:1981, Larichev:1995},  feeding energy at the latitude of the jet in a slowly propagating Rossby wave. The mechanism, which has been verified in  observations \citep{Lau:1988,Kug:2009}, simplified models \citep{Vautard:1988,Cai:1990,Robinson:1991}  and full GCMs \citep{Branstator:1992}, relies on the organisation of high frequency transients by the low frequency wave in such a way that vorticity fluxes due to wave-wave interactions provide a positive feedback to the low frequency flow itself.  In a QG model \cite{Cai:1990} showed that low frequency variability was entirely maintained by inverse energy cascade, while, when more realistic GCMs and observations are considered, its energetic contribution appears weaker than that provided by baroclinic and barotropic instability  \citep{Sheng:1990,Sheng:1990a}. 

Schneider et al.~\citep{Schneider:2004,Schneider:2006} pointed out that  in dry primitive equations models the heat vertically redistributed by the baroclinic eddies can reequilibrate the system by lifting the tropopause so that the atmosphere remains in a state of weak eddy-eddy interaction where the inverse energy cascade becomes inhibited. The two level primitive equations model of  HH  is likely to behave as a QG one, because the ``tropopause'' coincides with the top of the model and cannot be adjusted by the dynamics. Moreover GCMs feature an e--folding damping time due to horizontal diffusion in the free troposphere orders of magnitude bigger than in HH model, and cumulus convection schemes that can provide vorticity sources interacting with the low frequency wave \citep{Sardeshmukh:1988}. Therefore the dynamics and the energetics of the quasi--stationary waves in an aquaplanet could require a different interpretation from what was proposed in HH. 

If the persistent wave five could be satisfactorily interpreted as a neutral Rossby wave an equivalent barotropic vertical structure would be expected and an external energy source would be needed to maintain it against dissipation \citep{Hoskins:1981, Held:2002}, while if it were an unstable baroclinic wave \citep{Eady:1949} a westward tilt with height of the geopotential and a phase speed equals to the zonal wind at a certain level of the basic flow would be rather expected. It is possible to frame simple models able to set in a common framework Rossby--like waves and unstable baroclinic waves. The \cite{Green:1960} model, which describes the linearized dynamics of a stratified rotating fluid with vertical shear, shows  that marginally stable waves, associated to the transition between the two regimes,  have a phase speed equal to the zonal surface wind speed, so that LFV may not be incompatible with a stationary baroclinic process. 

The paper will show that baroclinic energetics and kinematics typical of equivalent barotropic waves  are both needed in order to explain the features of the observed quasi--stationary wave. Its wavenumber is controlled, in agreement with the Green's theory, by the averaged baroclinicity of the system, so that shorter quasi--stationary waves are generated as the equator to pole SST difference is decreased. Moreover, the role of the coupling of the extratropical dynamics with phase-locked tropical convection will be  analysed in detail. 
 
The paper is organised in the following way. In section \ref{sec:model} the model and the set of forcing SST distributions are introduced. In section \ref{sec:description} the kinematic properties of wave five are described, while in section \ref{sec:convection}  the tropical--extratropical interaction is discussed. An analysis of the wave's  vertical structure and energetics is performed in section \ref{sec:cospectra}, and a subsequent physical interpretation is proposed.  In section \ref{sec:baroclinicity}, we further extend our analysis by varying the baroclinicity of the system through changes in the meridional SST temperature gradient. A summary is given in section \ref{sec:summary} and the main conclusions are presented in section \ref{sec:conclusions}.
 
 \section{Model and Experiments} \label{sec:model}

With the expression "aquaplanet model" we refer to a general circulation model of the atmosphere whose lower boundary is given by a swamp ocean (motionless and of infinite heat capacity). The incoming radiation and the prescribed SST field, which through bulk formulas determine the surface heat flux, are zonally symmetric and constitute the two, independent, forcings of the system.

The simulations have been performed with an aquaplanet set up of the ECHAM5 model \citep{Roeckner:2006} according to the parameters and suggestions given in the Aquaplanet Intercomparison Project (APE). The resolution T31L19 has been adopted, but sensitivity runs at T63L31 have been also performed and are briefly  discussed in section  \ref{sec:summary}. The model is run in its standard configuration except for the cloud cover that is determined by the diagnostic scheme of \cite{Lohmann:1996} 
 
In addition to a standard incoming radiation pattern, the model is forced by the following set of SST distributions:
\begin{equation*}
	 \mathrm{SST}(\lambda,\phi)=\left\{
 	\begin{array}{ l l}
 	T_e-\Delta_T \cdot \sin^{2}(\frac{3\phi}{2}) & -\frac{\pi}{3} < \phi < \frac{\pi}{3} \qquad 0<\lambda \leq 2\pi \\
 	T_e-\Delta_T &  \text{otherwise}
 	\end{array} \right. ,
\end{equation*} 
where $T_e$ is the equatorial SST, $\Delta_T$ is the equator to pole SST difference, and the control experiment of the APE project  is recovered for $T_e=\Delta_T=27^\circ$C. The SST distribution is zonally and hemispherically symmetric and the prescribed temperature is constant on latitudes poleward of $60^\circ$. 
 
We have performed and analysed simulations with $T_e=\Delta_T=27^\circ$C (sections \ref{sec:description}--\ref{sec:cospectra}), and later examined the impact of varying $\Delta_T$  in $2^\circ$C steps from $27^\circ$C to $5^\circ$C (section \ref{sec:baroclinicity}). The global mean temperature, which is increased by decreasing $\Delta_T$, is a quantity of secondary relevance for the topic addressed in the paper, so that the choice of fixing the equatorial SST is justified. Moreover this choice makes basic features of the tropical atmosphere, as the lapse rate and the mid tropospheric temperature, less affected by changes in $\Delta_T$. The effect of changing the convective scheme and $T_e$ is  briefly discussed in section \ref{sec:summary}.

Ten years of 6-hourly sampled data have been gathered after one year of spin-up, which allows for a sufficiently accurate estimation of the spectral power density of the atmospheric waves. The analysis is done on daily mean data and, due to the symmetry of the aquaplanet model, it is  separately performed on the specular latitudes of the two hemispheres and the averaged result is presented. 

\section{The quasi--stationary wavenumber five}\label{sec:description}

\begin{figure}[h]
\noindent\includegraphics[width=0.6\textwidth]{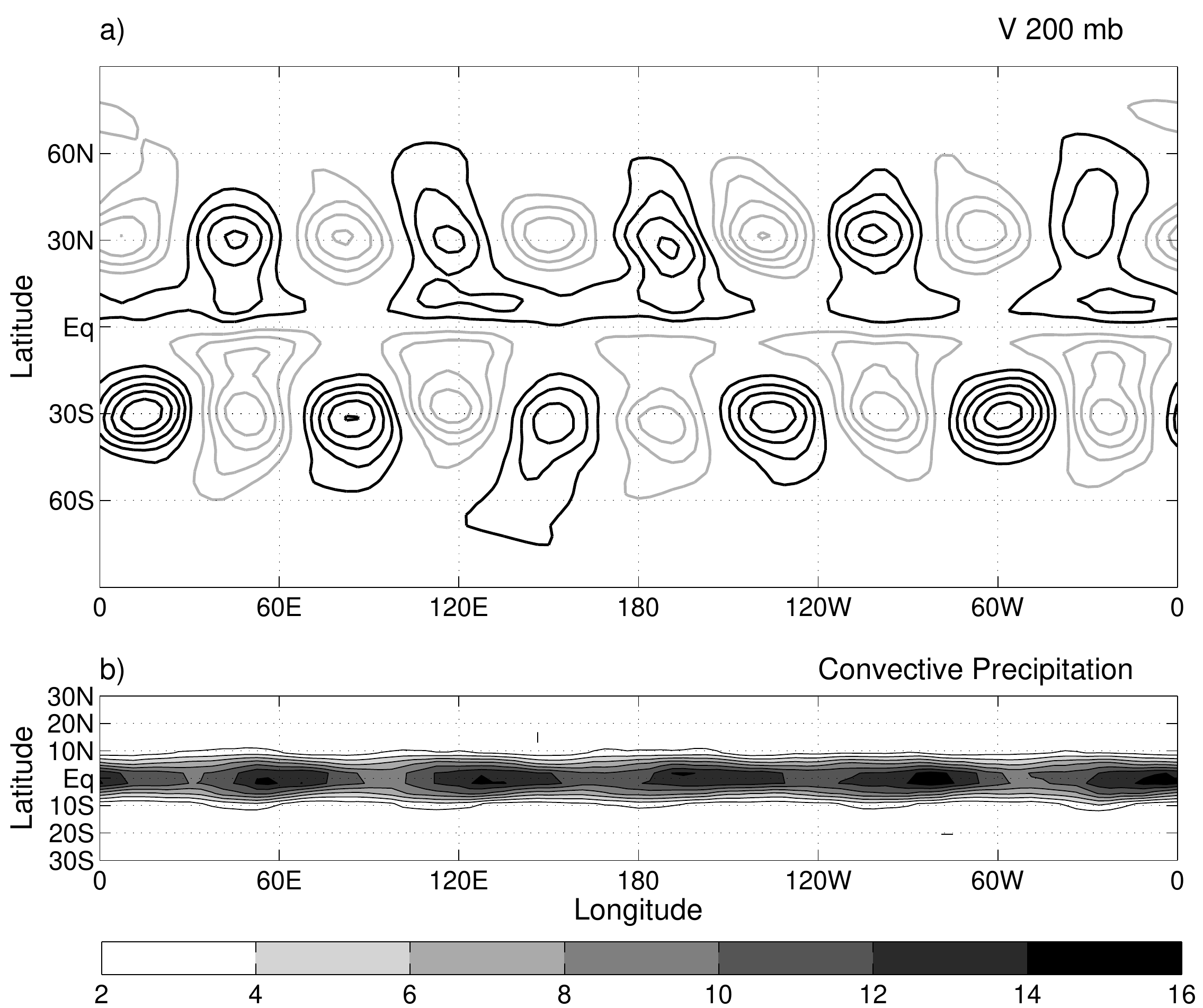}
\caption{A six months average of the (a) meridional velocity ($\mathrm{m\,s^{-1}}$) at 200mb and of the (b) convective precipitation (mm/day) in the tropical region calculated from an aquaplanet simulation setup according to the control experiment of the APE project  ($T_e=\Delta_T=27^\circ$C). In (a) c.i. is 2 $\mathrm{m\,s^{-1}}$ and   dark (grey) lines indicate positive (negative) values. Zero contour is omitted.  A low frequency wave five persists over the 6 months period in both the tropical and the extratropical motions. \label{fig:example_dt27}}
\end{figure}

The simulation performed with the setting of the control experiment of the APE project, shows a pattern dominated by the zonal component $k=5$, which persists over timescales much longer than the atmospheric internal low frequency variability \citep{Blackmon:1984}. To give a qualitative picture of the feature, Fig.~\ref{fig:example_dt27}a shows, as an example, a typical six months average of the meridional velocity at 200mb. A wave five of striking intensity peaks around $30^\circ$N/S and  it  extends over the  latitude band $20^\circ$--$50^\circ$. During the same time period, an analogous pattern is found in the tropical convective precipitation, featuring a deviation from the zonal mean of the order of 30\% (Fig.~\ref{fig:example_dt27}b).  The two hemispheres appear to be coupled with an upper tropospheric equatorial outflow (inflow) just west of  the longitudes where tropical precipitation is enhanced (reduced). Selecting a different 6 month time period would have given us an average field with a dominating wave five pattern with similar amplitude but, in general, different phase. 

Therefore, an ultra--low frequency wave five is present in both the tropics and the extratropics despite very different dynamical constraints governing the atmosphere at low and high latitudes. We have thus introduced two bidimensional fields that allow to separately analyse the wave motions in the tropical and in the extratropical regions:
\begin{align}
\hat{\Omega}(\lambda,t)=&<\omega(p_{\omega},\lambda,\phi,t)>_{0}^{15}\\
\hat{V}(\lambda,t)=&<v(p_{v},\lambda,\phi,t)>_{20}^{50},
\end{align}
where $\omega$ is the vertical velocity ($\mathrm{Pa\,s^{-1}}$), v the meridional velocity ($\mathrm{m\,s^{-1}}$), $p_{\omega}=500$mb and $p_{v}=200$mb the pressure levels at which the variables are respectively evaluated, and  $<{}>_{\phi_1}^{\phi_2}$ is an area weighted average between the latitudes $\phi_{1}$ and $\phi_{2}$. The latitudinal bands (both hemispheres are considered) and the levels have been chosen in order to contain the bulk of the low frequency power for all the set of simulations realised by varying $\Delta_T$ and $T_e$.

 \subsection{Wave's persistence}
 
 A quantitative evaluation of the persistence of a wave can be obtained by the plot of the mean amplitude of the zonal Fourier components of the wave averaged over different time windows. The calculation is performed by  partitioning the time domain in M non overlapping blocks of length $\tau$ days, and by averaging over all the blocks the amplitude of the mean wave component on each block. The averaging time ($\tau$) has been chosen equal to the powers of 2, ranging from 1 to 256 days, and the results plotted in a bi-log scale. Therefore, the most persistent component will appear as having the slowest average amplitude decrease as $\tau$ is increased.  
  
\begin{figure}[h]
\noindent\includegraphics[angle=0,width=0.5\textwidth]{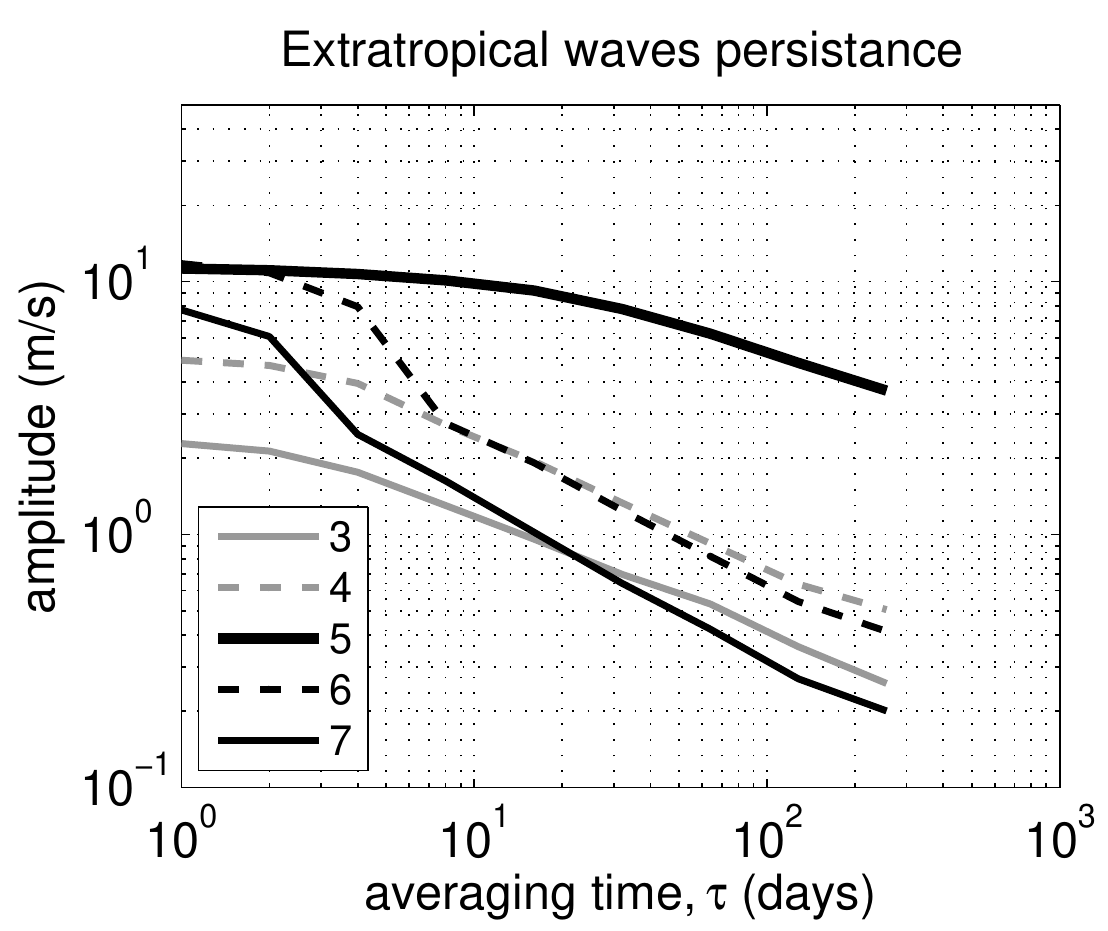}
\caption{Mean amplitude of the zonal Fourier components  in the extratropical 200mb meridional velocity ($\hat{V}$) averaged over as many non-overlapping time windows of $\tau$ days length. The amplitudes of zonal waves 3--7 are presented in a bi-log scale as a function of the averaging time $\tau$ itself. The wave $k=5$ features the smallest amplitude decrease with time, and it is therefore the most persistent wave.  \label{fig:amplitude}}
\end{figure}

This method has been applied to the zonal wave numbers 3--7 in $\hat{V}$, and results are shown in Fig.~\ref{fig:amplitude}.   The variance of the unfiltered (daily mean, $\tau=1$) $\hat{V}$ has an equal contribution from wavenumber five and six, but as the averaging time is increased the amplitude of the wave six drops down leaving the wave five notably stronger than all the other components on time scales longer than 10 days. A sharp drop in the wave amplitude over the synoptic timescales is visible for all the waves with $k>5$, while for $k\leq5$ a smoother decrease is observed, with the wave five featuring the smallest decrease for increasing $\tau$. When $\tau=32$ is considered, the amplitude of the wave five is $\sim70$\% of its $\tau=1$ value, whereas for $k=6$ the relative value realised for $\tau$=32 is just  10\% of the $\tau=1$ value. Such a strong persistence can be only realised if the wave five features very low frequency and high temporal coherence. The amplitude of $\hat{\Omega}$ as a function of $\tau$ leads to similar conclusions, and has not been plotted. 
  
\subsection{Spectra}\label{sec:spectrum}
The Hayashi spectra  \citep{Hayashi:1971} of $\hat{V}$ and of  $\hat{\Omega}$  have been calculated and results are presented in Fig.~\ref{fig:hayashi_v_om_dt27}. This technique, which is briefly described in the appendix, allows one to represent  the variance of a longitude--time field as a function of the zonal wavenumber and the frequency of the eastward and westward propagating waves composing it. Hayashi spectra have been used to analyse and compare atmospheric variability in GCM's and reanalysis \citep{Hayashi:1977a, Fraedrich:1978,Hayashi:1982,DellAquila:2005,Lucarini:2007,DellAquila:2007}.  Here, a two--sided representation has been adopted, in which the positive (negative) frequencies correspond to eastward (westward) propagating waves.

The power spectral density (PSD) of  $\hat{V}$ shows that clear peaks in the zonal wavenumbers 3--7 are organised along a non dispersive dispersion relation, which has been put in evidence  in Fig.~\ref{fig:hayashi_v_om_dt27}a by an ellipse enclosing the bulk of its spectral power.  At the reference latitude of  $30^\circ$ it corresponds to a group speed of about  $+40\,\mathrm{m\,s^{-1}}$. Specifically we observe a spectral peak centered at  $k=5$ and frequency close to zero, which is consistent with the picture of a quasi--stationary wave provided before.  Spectral peaks with $k<5$ ($k>5$) are westward (eastward) propagating. When inspecting the time evolution of the wave five phase (not shown), we discover that the wave alternates coherent periods of slow eastward and westward propagations which are the main causes of the slow loss of coherence of the wave presented in Fig.~\ref{fig:amplitude}. The distribution of spectral power on higher frequencies as the zonal wavenumber increases is a typical feature of extratropical waves spectra \citep{DellAquila:2005}, and it is usually interpreted as the signature of baroclinic unstable waves.  

As expected from the inspection of Fig.~\ref{fig:hayashi_v_om_dt27}b, a quasi--stationary wave five appears as a dominant feature also when considering the spectrum of $\hat{\Omega}$. Relevant low frequency spectral density is as well observed at wavenumbers $k=2$--$3$, which are typical of the Madden Julian Oscillation \citep{Madden:1994}, while  the dispersion relation on wave numbers 1--5,  underlined by the dotted line in Fig.~\ref{fig:hayashi_v_om_dt27}b, can be explained in terms of Kelvin waves propagating at the group speed of $\sim20\,\mathrm{m\,s^{-1}}$. This is in agreement with the speed observed in the actual climate system \citep{Kiladis:2009}.  As opposed to the extratropical case, spectral features for wavenumbers $k>5$ are barely present, consistently with the fact that baroclinic active waves are absent in the tropics. 

In order to test the presence of some coherence between the dominating patterns observed for the two fields described above, we estimate the Probability Distribution Function (PDF) of the phase difference between the wave five components in $\hat{\Omega}$ and in $\hat{V}$. The PDF that results is bell-shaped and peaks around zero. This clarifies the presence of a phase lock between the wave five in $\hat{\Omega}$ and in $\hat{V}$. For $k\neq 5$ the PDF is instead roughly flat. We can point to two possible distinct interpretations of the phase locking, which differ in the location of the forcing process:
\begin{itemize}
\item{The stationary extratropical wave is maintained by eminently local extratropical processes and induces 
a tropical convective pattern which acts as a positive feedback on the extratropical wave}
\item{The source of energy is eminently tropical, with the stationary convective pattern forcing the extratropical 
stationary wave by generating vorticity}
\end{itemize}
The experiment described in the next section is designed to identify the right picture of the process. 

\section{The role of tropical convection}\label{sec:convection}
In order to clarify the location of the wave's energy source, we perform an experiment where the non-zonal atmospheric forcing due to the tropical convection is suppressed and the zonal mean state of the atmosphere is not substantially altered. If the energy source were entirely in the tropics we would expect the low frequency mode to disappear, while if the tropical convection acts just as a feedback a change in the amplitude would be rather observed. 

This experiment is realised by zonally redistributing at every time step the tendencies produced by the convection scheme in the tropical region:
\begin{equation}
\frac{\partial \varphi}{\partial t}_{conv}^{R}=\alpha(\phi) {\left[\frac{\partial \varphi}{\partial t} \right]}_{conv}+(\alpha(\phi)-1)\frac{\partial \varphi}{\partial t}_{conv},
\end{equation}
where $\varphi$ represents a generic variable, namely q, u, v and T, \emph{conv} refers to the contribution to the tendency given by the mass flux convective scheme, the square brackets stand for a zonal average and $\alpha(\phi)$ is the degree of redistribution.  A complete zonal redistribution ($\alpha=1$) has been applied between $0^\circ$ and $15^\circ$ N/S, followed by a linear decrease to reach $\alpha=0$ at $20^\circ$ N/S, where the minimum in convective precipitation is located for all the experiments. The linear decrease has been introduced in order to reduce dynamical shocks due to discontinuities in the forcing.  It must be remarked that we have redistributed only the tendencies due to the parameterised moist convection, so that the dynamical tendencies are still locally determined at every longitude, and zonally asymmetric tropical motions are still generated, but are heavily reduced. The terms SC and CTL will be used to address the experiment with and without the symmetrized tropical convection, respectively. 

The difference in the zonal mean zonal wind  $[\bar{u}]$ between the SC and CTL experiments (contours), and  the  $[\bar{u}]$ of the CTL experiment (shaded) are plotted in Fig.~\ref{fig:compare_U}. We choose to compare the two experiments by using the $[\bar{u}]$ diagnostics because of its relevance as a parameter controlling Rossby wave propagation and the link of its vertical shear to the baroclinicity of the system.  Except for the upper equatorial region, the differences in $[\bar{u}]$ are smaller than $5\%$, and to a first order of approximation the mean state of the atmosphere remains reasonably unchanged by redistributing the convection. 

We now use the same analysis performed in Fig.~\ref{fig:amplitude} but to compare the amplitude of the low frequency waves  in $\hat{\Omega}$ and $\hat{V}$ between the two experiments. Fig.~\ref{fig:compare_wave} shows that in the SC experiment the tropical wave five is highly damped, while a persistent extratropical wave five still remains. Remembering that the vertical velocity is also a proxy for the forcing of extratropical Rossby waves by tropical divergent flow \citep{Sardeshmukh:1988}, it is concluded that an extratropical energy source must be maintaining the low frequency mode, and the first hypothesis proposed in section \ref{sec:description} is found to be appropriate.

The lower values in the extratropical wave amplitude for $\tau\sim15$ days in the SC experiment (Fig.~\ref{fig:compare_wave}b) is consistent with the picture of a positive feedback generated by the interaction with the tropical convection, which is confirmed by an analysis of the SC experiments over the whole set of $\Delta_T$ values (not shown). Moreover the synchronisation in the wave five activity between NH and SH (see fig.~\ref{fig:example_dt27}) is weaker in the SC experiments (not shown), so that the organisation of tropical convection on $k=5$ is one of the  mechanisms linking the two hemispheres.  The dynamics of the tropical--extratropical interaction and a detailed description of the feedback will be reported elsewhere.

\section{Energetics of the quasi--stationary wave}\label{sec:cospectra}

\subsection{Vertical wave's structure}
As confirmed in the previous section, the source of energy for the quasi--stationary extratropical wave 
has to be found in extratropical processes. Previous analyses have proposed that barotropic mechanisms should maintain the wave against dissipation \citep{Hendon:1985, Robinson:1991, Watanabe:2005}. Instead, we propose that baroclinic  energy conversion is responsible for feeding energy into the wave. We than test the very basic ingredient of baroclinicity, i.e. the presence of a phase tilt with height (baroclinic wedge) between the temperature and the meridional velocity of the wave. As is well known, in barotropic conditions, such a tilt is absent. 

The zonal tilt with height of the average phase of the meridional velocity and of the temperature in the low frequency wave five has been computed for the averaged wave in the $40^\circ$--$50^\circ$  (hereafter referred as region \emph{a}), $30^\circ$--$40^\circ$  (region \emph{b}) and $20^\circ$--$30^\circ$  (region \emph{c}) latitudinal bands. The calculation  has been performed as follows. At each pressure level ($\hat{p}$), the relative phase difference with respect to the reference level $p_{ref}=1000$mb is computed as:
\begin{equation}
	\gamma(\hat{p})=\arg<\phi_5(t,\hat{p}),\phi_5(t,p_{ref})>,
\end{equation}
where $\arg(z)=\arctan(\Im(z)/\Re(z))$ is the phase of the complex number $z$, $<\cdot,\cdot>$ stands for the correlation in time\footnote{The phase of the correlation between complex wave amplitudes --- also know as complex correlation --- is equivalent to the weighted time mean of the phase difference between the two waves, where the weight is the product of the respective amplitudes} and $\phi_5$ is the complex Fourier  amplitude of the component $k=5$ of a generic variable, namely v or T. The fields are preprocessed by applying a 10 days, 101 points long, Lanczos low pass filter \citep{Duchon:1979} in order to get rid of the high frequency fluctuations. The relative phase $\gamma_{rel}$ between the v and T wave is further obtained as:
\begin{equation}
	\gamma_{rel}=\arg<v_5(t,p_{ref}),T_5(t,p_{ref})>.
\end{equation}
We've verified that the results, which are plotted in Fig.~\ref{fig:tilt}, are not sensitive to the choice of the reference level.

We first observe that the v and T waves have opposite tilt, as is typical of baroclinic unstable waves. The baroclinic wedge is especially evident in region \emph{a}, where baroclinicity  is apparent throughout the troposphere, with vertical phase differences (about 1/12 of cycle, i.e 6 degrees) corresponding to about half of what is observed in the dominant Earth's baroclinic waves \citep{Lim:1991}.  As expected, in region \emph{c} the baroclinicity is lower since we are close to the tropical region, and almost all of the phase tilt for the v and T fields is located in the lowest levels.  Surprisingly, the baroclinicity  is even weaker in region \emph{b} which corresponds to the latitude band where the wave peaks:  this will be addressed in a later section. A more quantitative analysis of the baroclinic conversion processes requires estimating the various terms relevant for the Lorenz energy cycle \citep{Lorenz:1967}. Using a spectral approach, we have been able to highlight the contributions to the heat transport and to the energy conversion terms projecting on the various frequencies and wavenumbers.

\subsection{Heat transport}

The eddy contributions to the meridional heat transport $[\overline{vT}]$  at 750 mb and to the vertical heat transport $[\overline{-\omega T}]$ at 500mb  (where we have neglected $c_p$)  have been calculated by the cospectral technique introduced by \cite{Hayashi:1971} which is  reviewed in  appendix.  A meridional average in the $20^\circ$--$50^\circ$ latitude band has been preliminarily applied to v, T, and $\omega$. In Fig.~\ref{fig:cospectra_dt27} we show the positive values of the two cospectra $P_k^\nu(v,T)$ and $P_k^\nu(-\omega,T)$ using a logarithmic scale. The spectral components that feature negative values significantly different from zero at the 5\% confidence level  have been simply indicated by dots. Their values are however  at least an order  of magnitude smaller than the plotted positive ones. 

$P_k^\nu(v,T)$ and $P_k^\nu(-\omega,T)$ show an overall similarity on a large part of the spectrum, whose positive valued part  can be regarded as the region of active baroclinic waves. In fact, following the theory of the Lorenz's energy cycle, baroclinic unstable waves convert mean available potential energy into eddy potential energy, with a rate proportional to $P_k^\nu(v,T)$ times the meridional temperature gradient, whereas eddy potential energy  is converted into eddy kinetic energy, with a rate proportional to $P_k^\nu(-\omega,T)$ \citep{Lorenz:1967}. 

The analysis  will now focus on the region delimited by the continuous ellipse drawn in Fig.~\ref{fig:cospectra_dt27}a--b, which corresponds to the area where the spectral power in $\hat{V}$ is preferentially distributed (see Fig.~\ref{fig:hayashi_v_om_dt27}a). All the spectral peaks of the selected region with $k>5$ feature positive values in both $P_k^\nu(v,T)$ and $P_k^\nu(-\omega,T)$. This proves  rigorously their baroclinic nature. By contrast, the part of the region with $k<5$ features weak or negative energy conversion, so that the spectral peaks there seen in  Fig.~\ref{fig:hayashi_v_om_dt27}a are energetically equivalent to neutral Rossby waves. The low frequency wave five, which is found close to the neutral boundary separating unstable and neutral waves, is the longest wave of the region which converts energy via baroclinic processes. 

Outside of the main dispersion relation delimited by the continuous ellipse, a spectral region characterised by intense  meridional and vertical heat transport, which is delimited by the dotted ellipse drawn in  Fig.~\ref{fig:cospectra_dt27}a--b, is observed on k=2--5 and frequencies 0.05--0.2 $\mathrm{days^{-1}}$.  This \emph{secondary} dispersion relation, which does not dominate the PSD function of $\hat{V}$ shown in Fig.~\ref{fig:hayashi_v_om_dt27}a, corresponds to baroclinic active, shallow and meridionally narrow waves, as will be clarified by the arguments given in the subsection called \emph{interpretation}. 

To investigate the dependance on the latitude of the baroclinic energy conversion performed by the low frequency wave five, $P_k^\nu(-\omega,T)$ has been separately computed on  the three previously defined latitudinal bands. The three resulting spectra are shown in Fig.~\ref{fig:wt_10deg_band}a--c, and dotted circles have been drawn on the figures to indicate the spectral peak corresponding to the low frequency wave five.  While in region \emph{a} its energy  conversion is comparable to the one performed by the other spectral peaks, and the same applies for region \emph{c}, even if the baroclinic processes are overall much weaker, a reduced conversion is observed in region \emph{b}. In particular, the energy conversion is completely suppressed on the negative frequencies, which suggests that  baroclinic energy conversion is there realised only during periods of eastward propagation.  This latitudinal structure is in agreement with the tilts in the v wave observed in Fig.~\ref{fig:tilt}. A similar analysis for $P_k^\nu(v,T)$ (not shown) reveals that the low frequency wave five transports heat meridionally in all the latitude bands. Therefore, the wave five is an active baroclinic wave, but its energy conversion is concentrated on some latitudes and it is sensitive to the zonal phase speed of the wave.

Whether baroclinic conversion is really the leading process maintaining the wave will be clarified by the energy balance presented in the next subsection. 

\subsection{Spectral energy balance}

Barotropic instability and forcing by transient eddy fluxes are the atmospheric processes, other than the already mentioned direct baroclinic energy conversion, that could feed kinetic energy in the ultra low frequency wave five. The energetic contributions due to these process have been quantified by the cross--spectral method proposed by \cite{hayashi:1980}, which essentially provides a spectral picture of the Lorenz energy cycle. For each spectral component, this method estimates the linear energy transfers between the eddy kinetic and the potential energy reservoirs, which results from barotropic and baroclinic processes, and the redistribution of eddy kinetic energy among spectral components due to non linear wave interactions, which includes the low--frequency forcing by transient eddy fluxes. 

Using the \cite{hayashi:1980} notation, the kinetic energy's prognostic equation for the spectral component $(k,\nu)$, after integration over the whole atmosphere, can be written as:
\begin{equation}\label{eq:wv5_balance}
\frac{\partial K_k^\nu}{\partial t}=\underbrace{<K \cdot K >_k^\nu}_\text{wave--wave}+\underbrace{<K_0\cdot K_k^\nu>}_\text{wave--mean}+\underbrace{(-P_k^\nu(\alpha,\omega))}_\text{baroclinic}+D_k^\nu,
\end{equation}
where $K_k^\nu$ is the total kinetic energy on the spectral component ($k,\nu$). $<K \cdot K>_k^\nu$ stands for the redistribution of energy into spectral component ($k,\nu$) as a result of non linear interactions between different waves. In particular, the energy transfer occurs among triads of waves which are related in frequencies (and in wavenumbers) as $\nu$, $\mu$, $\nu \pm \mu$ ($k$, $l$, $k\pm l$).  $<K_0\cdot K_k^\nu>$ stands for the transfer of energy  between the wave of spectral component ($k,\nu$) and  the kinetic energy of the time mean flow by barotropic processes.  $<K \cdot K>_k^\nu$ and  $<K_0\cdot K_k^\nu>$ have been  respectively computed by the formulae  (3.12a--b) and (3.14a) of \cite{hayashi:1980}, and consist in the sum of terms involving cospectra between momentum and the convergence of momentum  (see \cite{hayashi:1980} for details). $\alpha$ is the specific volume and, as previously introduced, $P_k^\nu(\cdot,\cdot)$ indicates a 2D space--time cospectrum. Therefore, the term  $-P_k^\nu(\alpha,\omega)$ gives the spectral estimation of the direct baroclinic energy conversion performed by the spectral component ($k,\nu$). The last term refers to the kinetic energy lost on ($k,\nu$) by dissipative processes. The only difference with the method described in  \cite{hayashi:1980} is that 1D space or time cospectra have been substituted by 2D space--time cospectra computed as in Eq.~{\ref{eq:2D_cospectra}}.

We have directly computed the first three terms on the rhs of Eq.~\ref{eq:wv5_balance}. Global mean values are obtained by a vertical integration over eight pressure levels, ranging from 1000\,mb to 100\,mb, and a meridional area weighted average over all the latitudes of the model. Because we adopt a space--time decomposition, the lhs of Eq.~\ref{eq:wv5_balance} is zero, and dissipation can be estimated as the residual of the energy balance. The resulting spectral contributions to the energetics of the waves with $k=5$ are plotted in Fig.~\ref{fig:wv5_balance} as a function of frequency. Fig.~\ref{fig:wv5_balance} clearly reveals that baroclinic conversion is the dominant process feeding kinetic energy in the [-0.1,0.1] (cpd) frequency range, where the power of the low frequency  wave five is distributed. As typical of inverse energy cascade, high frequency transients, through the wave--wave interaction term, force positive kinetic energy in the low frequency band, but it represents just a minor contribution compared to the predominant baroclinic  energetics of the wave five. Barotropic wave--mean flow interactions are instead on average draining kinetic energy out of the wave five and are reinforcing the time mean jet. Representative values for the three processes, which have been obtained by integrating the  spectra over the [-0.1,0.1] frequency band, are reported in table \ref{tab:sources}.

\begin{table}[t]
\caption{Kinetic energy sources (+) and sinks (-) on the low frequency $k=5$ wave due to fundamental atmospheric processes.  Dissipation has been estimated as the residual closing the energy balance. The energetic contributions have been computed by integrating the unsmoothed energy transfer spectra over the  [-0.1,0.1] (cpd) frequency band.} \label{tab:sources}
\begin{center}
\begin{tabular}{c|c}
 Process &  Energetic contributions $\mathrm{(W\,m^{-2})}$ \\
 \hline
 Baroclinic   &  0.16    \\
 Wave--Wave  &  0.04   \\
 Wave--Mean &  -0.06 \\
 Dissipation*  & -0.14 
\end{tabular}
\end{center}
\end{table}

Another peak in baroclinic energy conversion is present at $\nu \sim 0.2$ cpd. This is due to a fast propagating wave five that belongs to the secondary dispersion relation already mentioned when discussing Fig.~\ref{fig:cospectra_dt27}. 

\subsection{Interpretation}\label{sec:interpretation}

These results lead us to introduce a new paradigm that  describes the extratropical quasi--stationary wave five as a marginally stable baroclinic wave, and to look for theoretical models to justify its low phase speed and the latitudinal dependance of the efficiency of the energy conversion.

\cite{Green:1960} analysed the linear stability properties of an infinitely meridional extended wave ($l=0$) when the beta effect is added to the Eady's model. Due to the long wave stabilization provided by the beta effect, his model contains both the neutral Rossby waves for $k<K_c$,  and the baroclinically unstable waves for $k>K_c$, where $K_c$ is the critical total wavenumber separating the two regimes. A marginally stable wave of phase speed equal to the surface zonal wind is found for $k=K_c$ (see Fig.~2 in \cite{Green:1960}), so that  quasi--stationarity and marginal stability are two properties that coexist in the same wave. Components with $k<K_c$ ($k>K_c$) are respectively westward (eastward) propagating. These features are in agreement with the previously described kinematic and energetic properties of the waves in the main dispersion relation around the marginal condition $k=5$.

The fixed ($l=0$) meridional wavenumber of the Green's model does not limit its interpretative value, because 
the waves in the main dispersion relation feature a meridional scale ($L_b$) close to the width of the jet stream, and thus nearly independent of $k$. The width of the baroclinic zone is indeed a favourite meridional scale for the most unstable wave \citep{Hoskins:1981a}, provided that the zonal wavenumber is sufficiently high to satisfy the necessary condition for baroclinic instability
\begin{equation} \label{eq:critical}
k^2+{l_b}^2>{K_c}^2,
\end{equation}
where $l_b$ is a meridional wavenumber associated to $L_b$.

The presence of the secondary streak of peaks in Fig.~\ref{fig:cospectra_dt27} (see the dotted ellipse) can also be explained in terms of linearized baroclinic models. As $k$ is decreased under five, relation \ref{eq:critical} is no longer satisfied, and the meridional scale of the most unstable baroclinic wave has to become smaller than $L_b$ to allow further baroclinic conversion. As a result, the secondary dispersion relation can be interpreted in terms of active baroclinic waves that are zonally longer and meridionally narrower respect to the active waves in the main dispersion relation \citep{Hoskins:1981a}.
 
In Fig.~\ref{fig:wt_10deg_band} we have seen that the  energy conversion by the low frequency  wave five has a peculiar latitudinal structure with relevant inhibition in the $30^\circ$--$40^\circ$ latitudinal band. An explanation is provided considering the bounds on the zonal phase speed ($c_r$) of unstable baroclinic waves calculated by \cite{Pedlosky:1979} in QG $\beta$ channel conditions:
\begin{equation}\label{ped}
U_{min}-\frac{\beta}{2(\frac{\pi^2}{4L^2}+k^2)}\leq c_r \leq U_{max},
\end{equation}
where $U_{min}$ and $U_{max}$ are respectively the minimum and maximum of the zonal wind in the meridional plane, and L the width of the channel. Due to the $\beta$ effect, relation $\ref{ped}$  states that  in the presence of a westerly vertical shear unstable waves can propagate even slower than the minimum zonal surface wind. If a channel centered at $35^\circ$ of latitude and $30^\circ$ wide is taken as representative of the extratropical baroclinic zone of the aquaplanet, the correction\footnote{Variations of ten degrees in the channel's parameters lead to changes in the correction due to the $\beta$ effect $<1$ $\mathrm{m\,s^{-1}}$ }  due to the $\beta$ effect for $k=5$ is of the order of 6 $\mathrm{m\,s^{-1}}$.  Subtracting this value from the zonal mean zonal wind at the lowest model level, which is shown in Fig.~\ref{fig:ll_wind}, reveals that a quasi--stationary wave five can be unstable thanks to the presence of the weak westerly  winds on the northern and southern flank of the surface jet stream, which is the latitude range where baroclinic conversion indeed occurs. On the contrary, in the  $30^\circ$--$40^\circ$ latitude band the lower bound on the phase speed implies that waves with frequency $\lesssim +0.025$ should be stable, which is consistent with  the weak  conversion observed in Fig.~\ref{fig:wt_10deg_band}b.

The weak baroclinic structure of the wave five in the $30^\circ$--$40^\circ$ latitude band suggests that its kinematic could be explained in terms of barotropic Rossby wave propagation. This is coherent with the previous energetic interpretation based on the Green's model, because marginally stable waves constitute the transition between Rossby and baroclinic active wave regimes. Both  the Rossby and the Eady models are essentially contained into Green's. The  \emph{sharp} 200 mb jet, which can be seen in Fig.~\ref{fig:compare_U}, creates an upper tropospheric wave guide  where a stationary Rossby wave five can be channeled  \citep{Hoskins:1993}. This is demonstrated in Fig.~\ref{fig:ks_rossby_dt27}, where the stationary Rossby wave number at 200 mb \citep{Hoskins:1981} is plotted against latitude. The channel delimited by the latitudes at which $k=5$ is stationary is centered at $30^\circ$,  in qualitative agreement with the latitude where the amplitude of the wave five peaks. This interpretation may be improved noting that the wave is more meridionally extended than the channel ($\sim30^\circ$ for the wave, against $\sim15^\circ$ for the channel), so that the wave five may rather see the climatological mean jet as a PV jump. Under this conditions, Rossby waves are trapped and their propagation non dispersive \citep{Schwierz:2004}, in agreement with the observed main dispersion relation. 

\section{Sensitivity to the baroclinicity of the system}\label{sec:baroclinicity}

Changing the equator to pole SST difference ($\Delta_T$), as presented in section \ref{sec:model}, amounts to altering the mean baroclinicity of the system. Analysing how the properties of the low frequency variability of the system depend on the temperature gradient allows us to test the robustness of our interpretation based on the Green's model.

In Fig.~\ref{fig:amplitude_dt}  we plot the mean amplitude of $\hat{V}$ for the zonal waves five, six and seven at the averaging time of 1 day (left) and 32 days (right) as a function of $\Delta_T$. The two timescales have been chosen to measure the total and the low frequency mean wave amplitude respectively. As $\Delta_T$ is decreased the total wave amplitude decreases, in agreement with the reduced baroclinicity of the system \citep{Stone:1978}, while the dominant low frequency mode drifts toward higher wave numbers. Wave five, six and seven reach their optimal  stationary condition for $\Delta_T=27$, $\Delta_T=15$ and $\Delta_T=11$ degrees respectively. This shift is in agreement with the displacement of the Green's model marginally stable baroclinic wave toward higher wavenumbers as the vertical shear of the basic state is decreased. 

The stabilized waves six and seven feature a  structure closely resembling the one already described for the wave five. As an example, Fig.~\ref{fig:example_dt15} shows six months average of the experiment $\Delta_T=15$ (compare it with the case $\Delta_T=27$ shown in Fig.~\ref{fig:example_dt27}). An equatorward displacement of $\sim 5^\circ$ of the peak in the extratropical V wave, linked to the displacement of the jet stream itself, and a different tropical convective pattern of precipitation, are the only two remarkable differences of an otherwise similar pattern projecting on the zonal wavenumber six instead of five. SC experiments verified that the tropical convection is again responding and not forcing the extratropical wave. 

The cospectra $P_k^\nu(v,T)$ and $P_k^\nu(-\omega,T)$ are shown in Fig.~\ref{fig:cospectra} for the selected simulations $\Delta_T=21$, $\Delta_T=17$ and $\Delta_T=11$. The waves considered in the following discussion are those identified by the spectral power contained in the  main dispersion relation of each experiment, which have been for clarity indicated in Fig.~\ref{fig:cospectra} by ellipses. For $\Delta_T=21$ the wave five is still both the dominant low frequency wave and an active baroclinic wave vertically and meridionally transporting heat, but  as $\Delta_T$ decreases its energy conversion is reduced and gets completely suppressed for $\Delta_T=17$. The wave five has become a neutral westward propagating Rossby wave, which is only meridionally transporting heat and it is mainly forced by  inverse energy cascade. But at this stage the wave six, which is both meridionally and vertically transporting heat,  is the dominant low frequency wave, and the new marginally stable baroclinic component instead of wave five. By further decreasing $\Delta_T$ also the wave six becomes neutral, and the wave seven takes its place as the marginally stable quasi--stationary baroclinic component (Fig.~\ref{fig:cospectra}e-f).

In all the experiments, a weak but significant counter baroclinic energy conversion is observed on the westward propagating components of   the main dispersion relation (see Fig.~\ref{fig:cospectra}b,d,f and in Fig.~\ref{fig:cospectra_dt27}b). This inversion of the Lorenz energy cycle may not be surprising, because those components, which are baroclinically neutral,  are principally forced by inverse energy cascade of kinetic energy, so that they have a source of kinetic energy but miss a source of potential energy. Therefore, a counter baroclinic energy conversion is necessary to produce the eddy available potential energy required to maintain thermal wind balance.

\section{Summary}\label{sec:summary}

This paper has been devoted to studying the statistical properties of the ultra--low frequency variability observed in many aquaplanet models and to identifying the physical mechanisms responsible for its maintenance. Past experiments performed with a variety of atmospheric models have shown that when zonally symmetric boundary conditions are considered for TOA incoming radiation and swamp ocean SST fields the extratropical atmospheric variability features an extraordinary strong signature of quasi--stationary waves featuring impressive temporal coherence well beyond the typical atmospheric time scales. In the parametric range roughly corresponding to the present climate --- as that considered in the Aquaplanet Intercomparison Project (APE) --- the ultra LFV manifests itself in the form of a wave with characteristic wavenumber five. 

A previous study \citep{Hendon:1985}, based on a simplified two level primitive equation model, explained this feature using the paradigm of a Rossby wave forced by a turbulent inverse energy cascade. Nevertheless the properties of the wave five have never been analysed in detail in more complete GCMs, even though the topic is of interest considering the resemblance with some quasi--stationary wave patterns observed in the Earth's atmosphere.

In the simulations we have performed with an aquaplanet set up of the ECHAM5 model, using boundary conditions analogous to those in the APE project, the wave five peaks in the subtropics, where it has a weak baroclinic vertical structure, it extends in the $20^\circ$--$50^\circ$ latitudinal band and it features a very high temporal persistence. A wave five pattern in the tropical convection is found to be phase--locked with the extratropical wave, so that an experiment in which the forcing of extratropical waves by the tropical convection is suppressed has been realised to investigate the nature of the  interaction. It has been shown that the tropics are just responding as a positive feedback, and a prevalent extratropical dynamics  maintains the wave five. 

After performing a spectral analysis of the meridional velocity field, and of the meridional and vertical heat transports,  the wave five results as a spectral peak of nearly zero frequency belonging to a well defined dispersion relation on the zonal wave numbers $k=3$--$7$, which corresponds to a non dispersive wave packet propagating eastward at about 40 $\mathrm{m\,s^{-1}}$. Moreover, the wave five is the longest component on the dispersion relation which is converting available into kinetic energy through baroclinic processes.  The Green's model of baroclinic instability has been found of great guidance in evaluating these results, and has lead us to introduce a new paradigm for describing the low frequency wave five as a marginally stable baroclinic wave. A spectral analysis of the kinetic energy balance on $k=5$ reveals that inverse barotropic energy cascade provides kinetic energy to the low frequency wave, but this is about just one forth of the energy converted by baroclinic processes. 

The Green's model, and Pedlosky's bounds on the phase speed of unstable waves in QG $\beta$ channel conditions, show that marginally stable waves feature a zonal  phase speed close to the zonal surface wind speed, so that the low level mean wind becomes one of the main parameters controlling the stability of the wave. Therefore, in the $30^\circ$--$40^\circ$ degrees latitude band, where the mean surface jet peaks,  the wave five features a limited energy conversion and it propagates as an almost barotropic equivalent wave meridionally trapped by the upper tropospheric PV jump created by the jet stream. Baroclinic conversion is instead enhanced on the southern and northern flank of the jet, where low level winds are closer to zero, and baroclinic energy conversion of a quasi--stationary wave becomes possible.

The proposed interpretation has been tested in a set of experiments in which the baroclinicity of the system has been changed by decreasing the equator to pole SST difference ($\Delta_T$). In this previously unexplored parametric range, quasi--stationary waves of wavenumber six and seven, featuring a structure analogous to the wave five, are respectively found for $\Delta_T=15$ and $\Delta_T=11$.  This shift in the wave number is consistent with the displacement of the marginally stable wave of the Green's model as the vertical shear is decreased, so that the interpretation is found to hold over the whole parametric range.

Other experiments (not discussed) proved that the selection of the stationary zonal wave number by $\Delta_T$, and the interpretation of the LFV as a marginally stable wave, is robust  against changes in the equatorial SST, in the closure of the convective scheme and in the model's resolution. The organisation of the tropical convection on the quasi--stationary wavenumber is instead  highly sensitive to changes  in the resolution and in the convective scheme. Nonetheless, the pattern remains visible in monthly mean averages for those values of $\Delta_T$ associated to optimal stationary conditions. 
  
\section{Conclusions and Discussion}\label{sec:conclusions}
The main findings of this paper are:
\begin{itemize}
\item{Aquaplanet models setup according to the APE project  feature strong quasi--stationary waves trapped along the jet stream, which are mainly maintained against dissipation by direct baroclinic energy conversion. The forcing by high frequency transients, which had been previously observed by \cite{Watanabe:2005}, only provides a positive feedback to the wave}
\item{The Hayashi spectral analysis of the energetics of the extratropical waves is a powerful technique to identify the stability of the different waves and their organisation along preferential dispersion relations.  The theoretical dispersion relation linking Rossby to baroclinic unstable waves has been easily identified by this technique in the aquaplanet. }
\item{The Green's model has been found of valid guidance to interpret the dispersion relation and to identify the quasi--stationary waves as marginally stable baroclinic waves. As far as we know, it is the first time that the transition between long Rossby waves, and short unstable baroclinic waves, passing through the low frequency marginally stable wave, has been observed in a non linear GCM.}
\item{According to the Green's model, the wavenumber of the stationary component is controlled by the average baroclinicity of the system, which can be changed by manipulating the equator to pole SST difference. Barotropic reasoning, based on the Rossby model, could explain the right kinematics and stationary wavenumber, but it would never be able to represent the right energetics which needs to consider the baroclinic nature of the wave.}
\item{ The marginally stable baroclinic waves organise the tropical convection so that it provides a positive feedback on the baroclinic wave itself. The classical framework of anomalous tropical convection forcing stationary Rossby waves is therefore here reversed.}
\end{itemize}

This perspective on LFV generation is not necessarily exclusive of aquaplanet models, but may be appropriate every time a marginally stable baroclinic wave and an upper tropospheric wave guide are present. Our interpretation of the aquaplanet's wave five is indeed comparable to the leaky normal mode theory proposed by \cite{Salby:1982} to explain the SH DJF wave five mode, whose properties were found by the author as  ``not inconsistent with features of baroclinically unstable mode''.  In particular the nearly barotropic phase structure in the amplified region which becomes more baroclinic on the two sides is a remarkably similar characteristic between the two waves. 

 Even though the zonal asymmetries of the NH boundary conditions make a direct comparison with the circumglobal low frequency patterns individuated by \cite{Branstator:2002} and \cite{Ding:2005} harder, the participation of a marginally stable wave in their energetics should not be excluded and may be worth further research.  Particularly interesting are those cases, such as the NH summer 2007 \citep{Blackburn:2008}, where a dynamical similarity with aquaplanets models is realised because a closed jet stream pattern is observed as a result of a negative NAO pattern. 
 
The misrepresentation of extratropical quasi--stationary baroclinic processes could also contribute to the systematic error of climate models. In particular, it might help to explain the big differences in meridional enthalpy transport  observed  between state of the art climate models  in the SH \citep{Lucarini:2010}.
  
Further investigations will deal with quantitatively comparing the dispersion relation observed on  the aquaplanet with that predicted by the linearized model of baroclinic instability, and with  computing  the normal modes of the time mean state.  Moreover  the dynamics of the LFV will be further investigated with respect to the feedback of tropical convection, the interaction with the high frequency and the sensitivity to the zonal symmetry of the jet stream. This knowledge will clarify the extent to which the interpretation of LFV as marginally stable baroclinic waves can be relevant in the Earth's atmospheric dynamics.

%%%%%%%%%%%%%%%%%%%%%%%%%%%%%%%%%%%%%%%%%%%%%%%%%%%%%%%%%%%%%%%%%%%%%%
%% FIGURES
%%%%%%%%%%%%%%%%%%%%%%%%%%%%%%%%%%%%%%%%%%%%%%%%%%%%%%%%%%%%%%%%%%%%%%

\newpage

\newpage

\newpage

\begin{figure}[h]
\noindent\includegraphics[width=\textwidth]{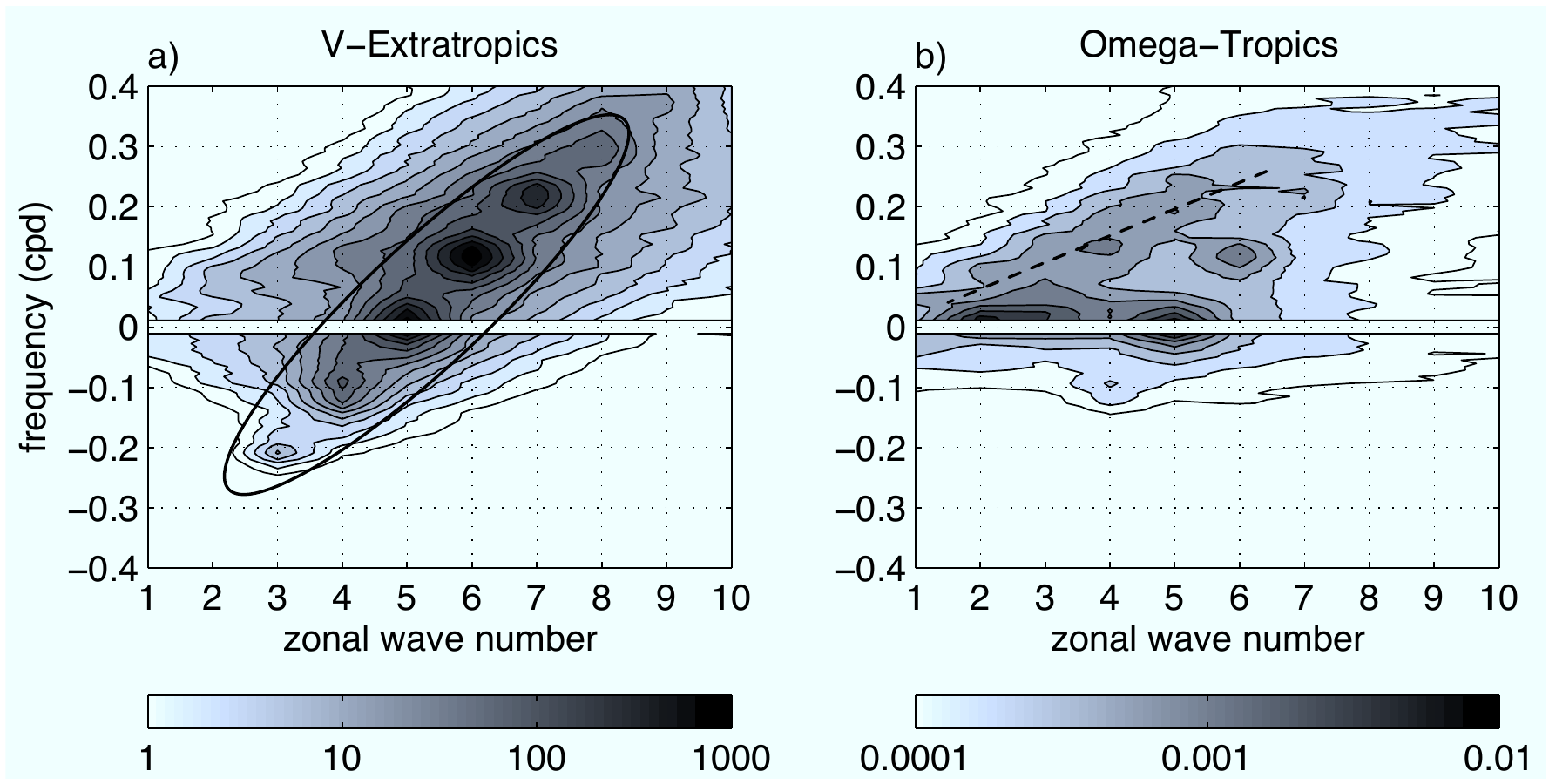}
\caption{Hayashi power spectrum of (a) the 200mb meridional velocity averaged in the extratropics ($\hat{V}$)  and (b) of the 500mb vertical velocity averaged in the tropics ($\hat{\Omega}$) for the control experiment. The spectrum is represented in a linear and two--sided representation, where frequencies are expressed in cycles per days (cpd), with positive (negative) values corresponding to eastward (westward) propagating waves respectively. The spectral density is presented in a logarithmic scale, and four contours are plotted every order of magnitude. Units are in $\mathrm{m^2\,s^{-2}\,day}$ (a) and  $\mathrm{Pa^2\,s^{-2}\,day}$ (b). The ellipse in (a) indicates the main spectral region where extratropical wave activity is distributed, while the dotted line in (b) indicates the dispersion relation of tropical kelvin waves. \label{fig:hayashi_v_om_dt27}}
\end{figure}

\newpage

\begin{figure}[h]
\noindent\includegraphics[width=0.5\textwidth]{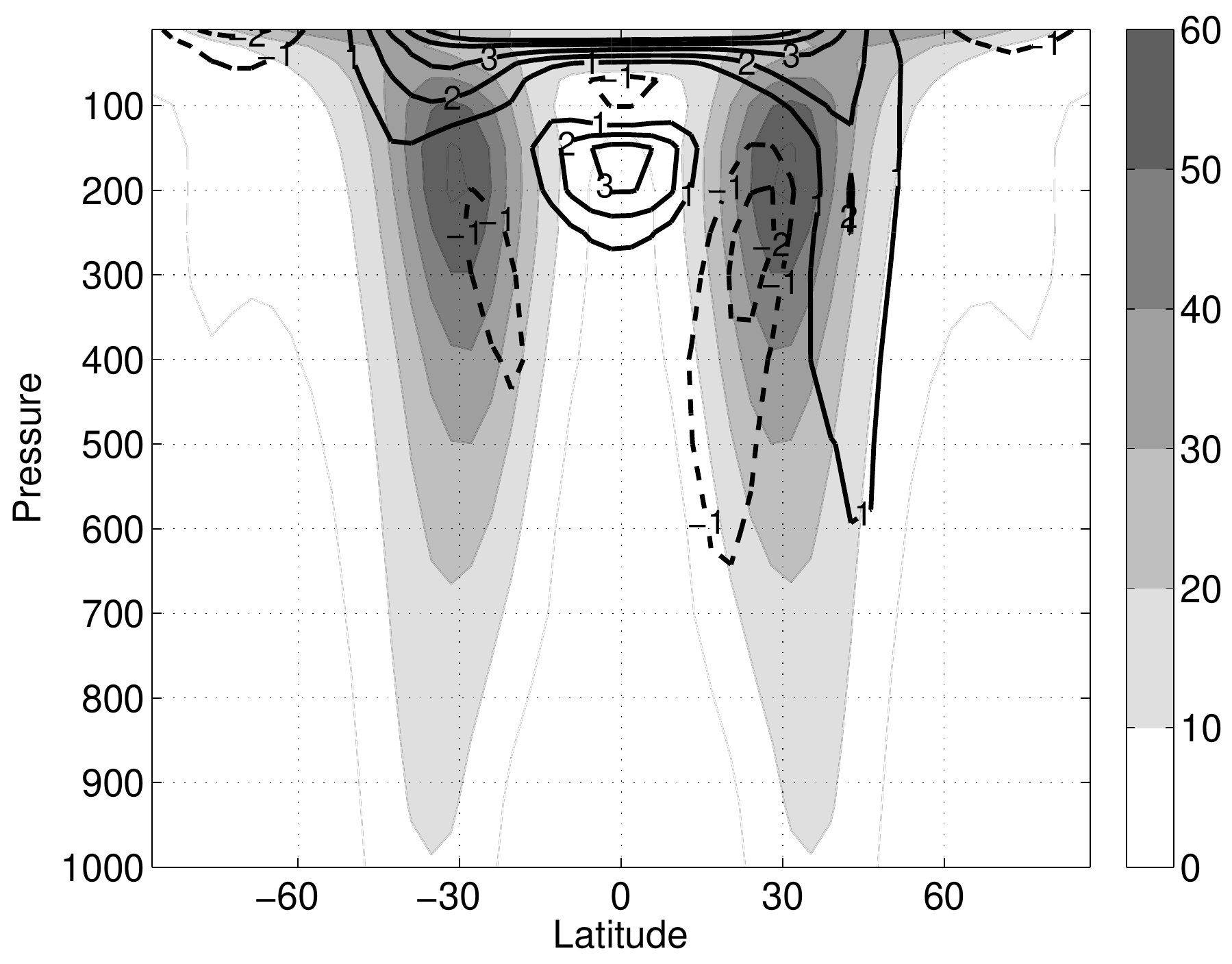}
\caption{Zonal mean of the time mean zonal wind ($[\bar{u}_{CTL}]$) in the control (CTL)  experiment (shaded) and difference in the same field  between the zonally redistributed tropical convection (SC) experiment and the control ($[\bar{u}_{SC}]-[\bar{u}_{CTL}]$) (contour). Continuous (dashed) lines indicate positive (negative) values, and c.i. is 1 $\mathrm{m\,s^{-1}}$. The difference in the mean state between the two experiments in the extratropics is generally smaller than the 5\%. \label{fig:compare_U}}
\end{figure}

\begin{figure}[t]
\noindent\includegraphics[width=0.9\textwidth]{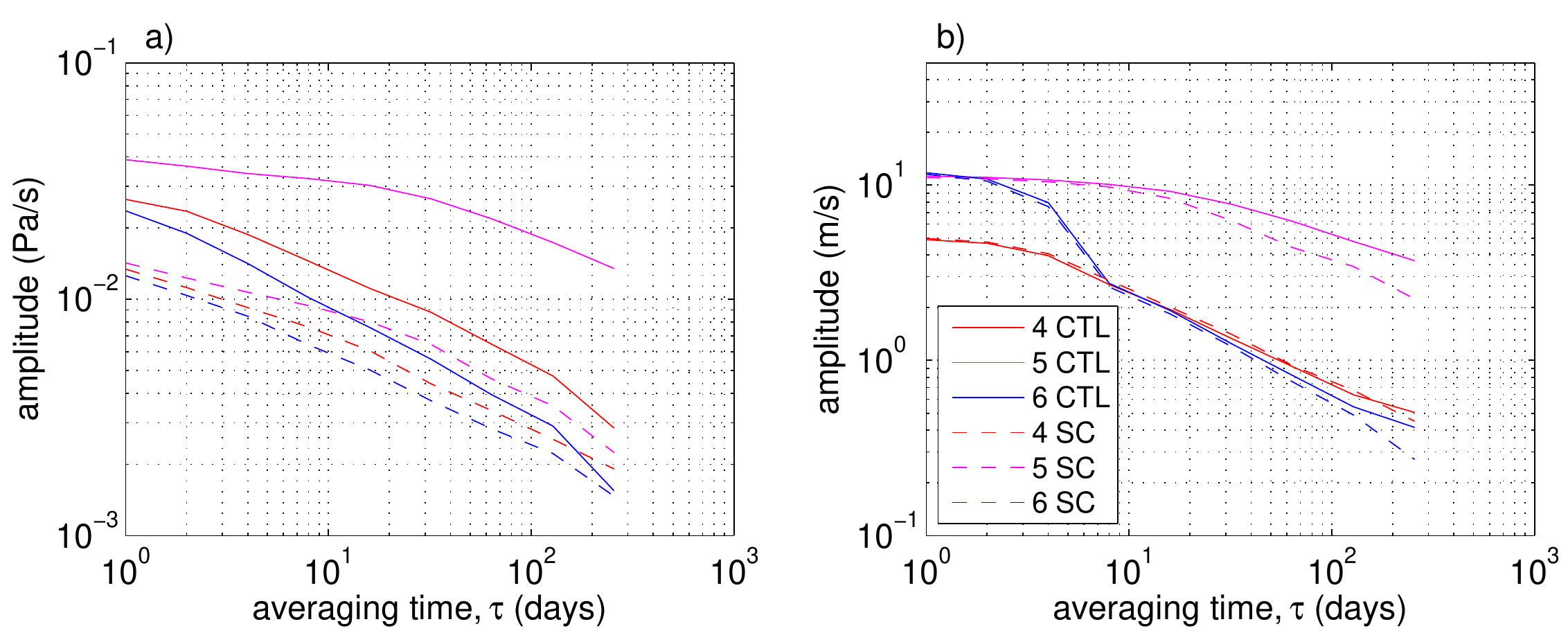}
\caption{As in Fig.~\ref{fig:amplitude} but to compare the amplitude of the waves as a function of the averaging time between the control (CTL, full line) and  the zonally redistributed tropical convection (SC, dotted lines)  experiments. The comparison is performed on (a) the tropical vertical velocity ($\hat{\Omega}$) and on (b) the extratropical meridional velocity ($\hat{V}$). The symmetrization of convection inhibits the wave five in the tropical convection while it just weakens its persistence in the extratropics.  \label{fig:compare_wave} }
\end{figure}

\begin{figure}[t]
\noindent\includegraphics[width=0.9\textwidth]{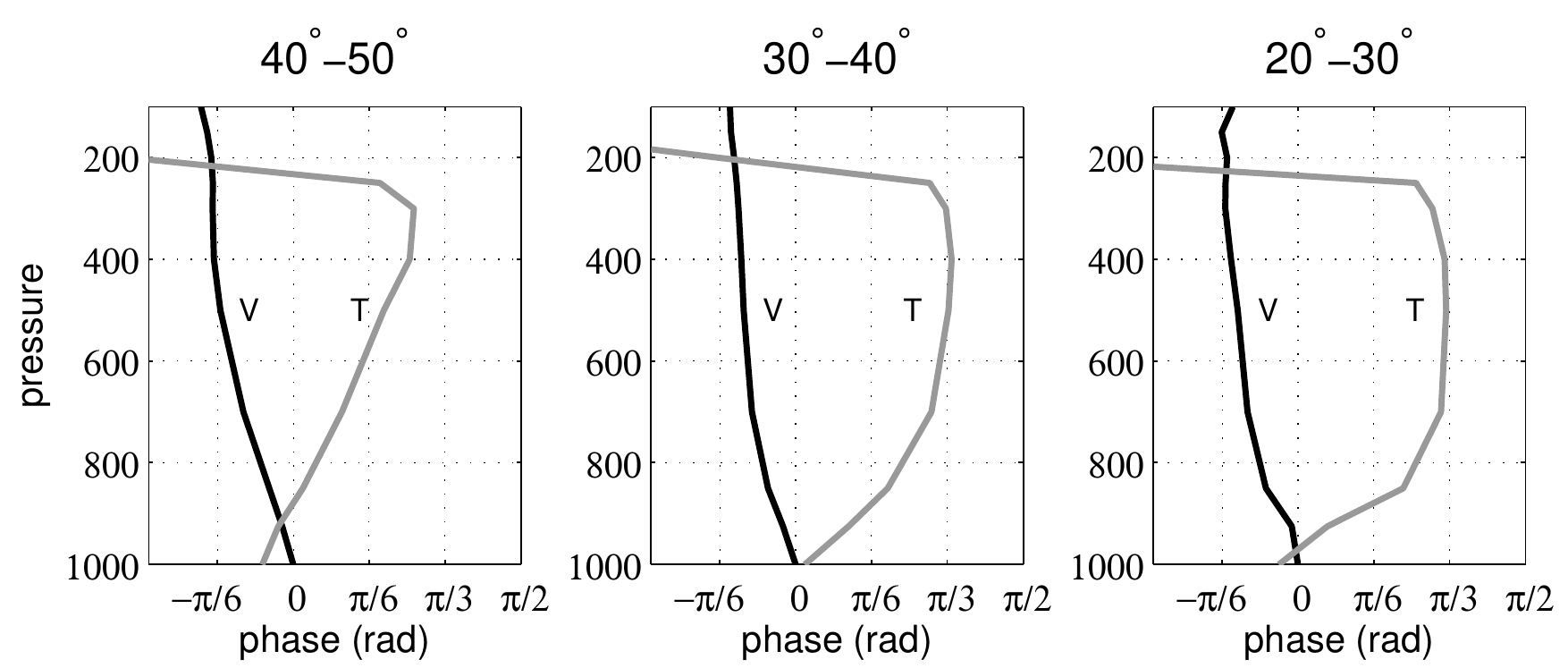}
\caption{Mean zonal tilt with the height of the meridional velocity and temperature fields in the low frequency wave five. The vertical structure is separately calculated for the meridionally  averaged wave on the latitude bands (a) $40^\circ$--$50^\circ$, (b)  $30^\circ$--$40^\circ$ and (c) $20^\circ$--$30^\circ$.  \label{fig:tilt}}
\end{figure}

\begin{figure}[t]
\noindent\includegraphics[width=\textwidth]{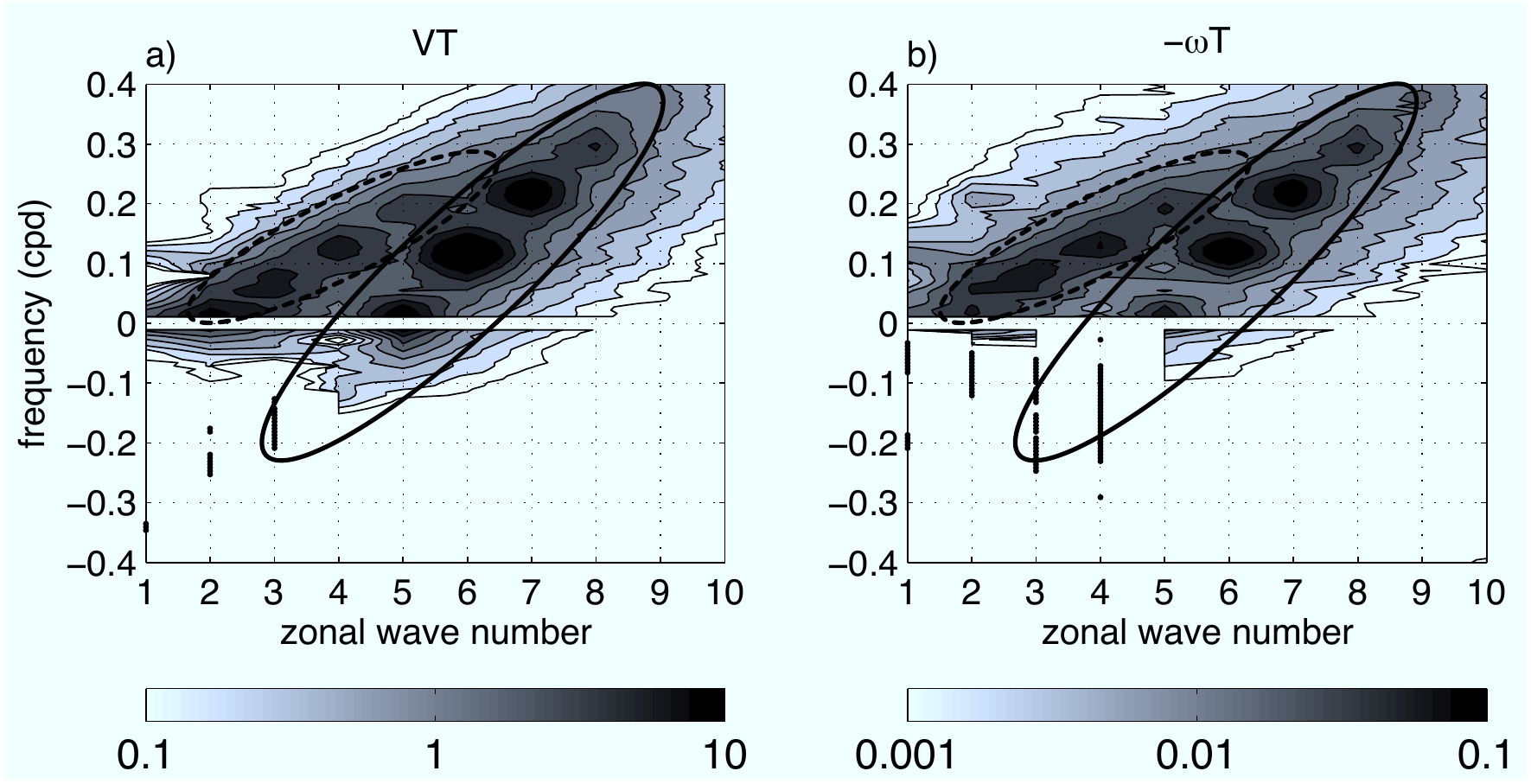}
\caption{Spectral representation of (a) the meridional heat transport at 750mb and of (b) the vertical heat transport at 500mb as calculated by an Hayashi cospectral technique. Units are in $\mathrm{m\,s^{-1}\,K\,day}$ (a) and $\mathrm{Pa\,s^{-1}\,K\,day}$ (b). The specific heat $c_p$ has been neglected and a meridional average in the $20^\circ$--$50^\circ$ latitude band, where the wave five peaks,  has been previously applied to the data. The plotting conventions are the same as those in Fig.~\ref{fig:hayashi_v_om_dt27}, and the dotted areas correspond to regions featuring negative values significantly different from zero. The continuous and the dotted ellipse indicate the main and the secondary extratropical dispersion relations,  respectively.  \label{fig:cospectra_dt27}}
\end{figure}

\begin{figure}[t]
\noindent\includegraphics[width=0.4\textwidth]{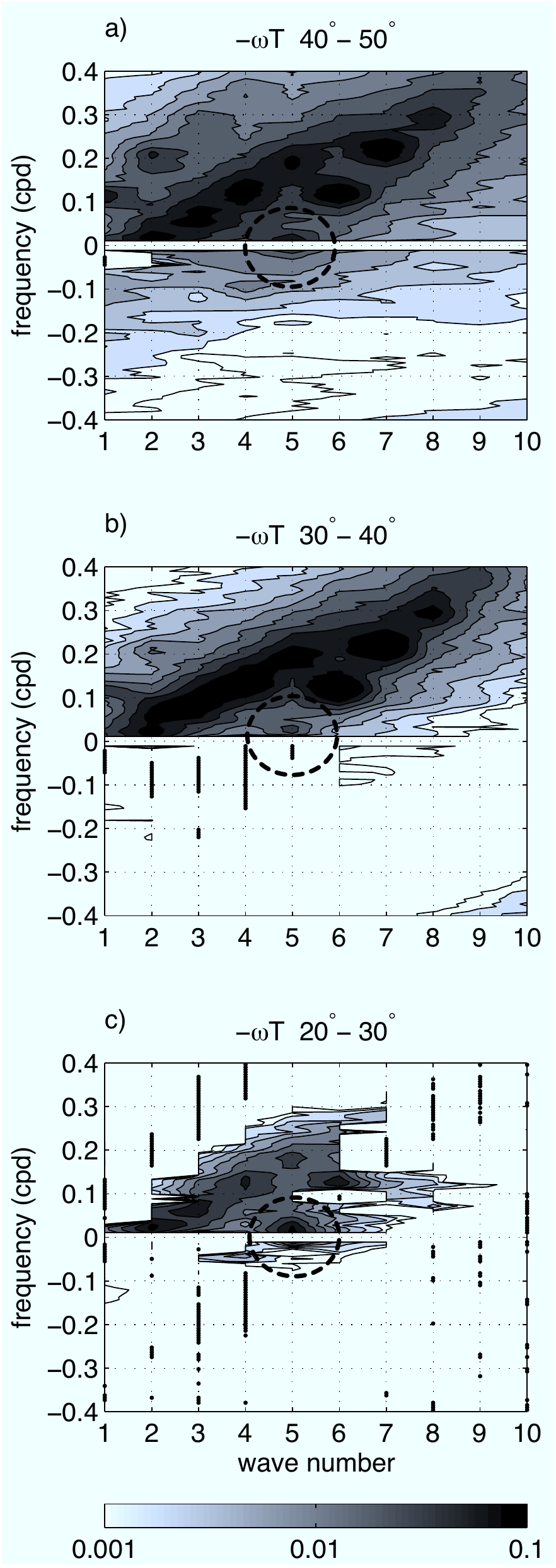}
\caption{As in Fig.~\ref{fig:cospectra_dt27}b but for the the vertical heat transport at 500mb  due to the meridionally averaged waves in the (a) $40^\circ$--$50^\circ$ , (b) $30^\circ$--$40^\circ$ and (c) $20^\circ$--$30^\circ$ latitude bands. The dotted circle points to the  baroclinic energy conversion by the low frequency wave five. Note the inhibition in the central latitude band.  \label{fig:wt_10deg_band}}
\end{figure}

\begin{figure}[t]
\noindent\includegraphics[width=0.5\textwidth]{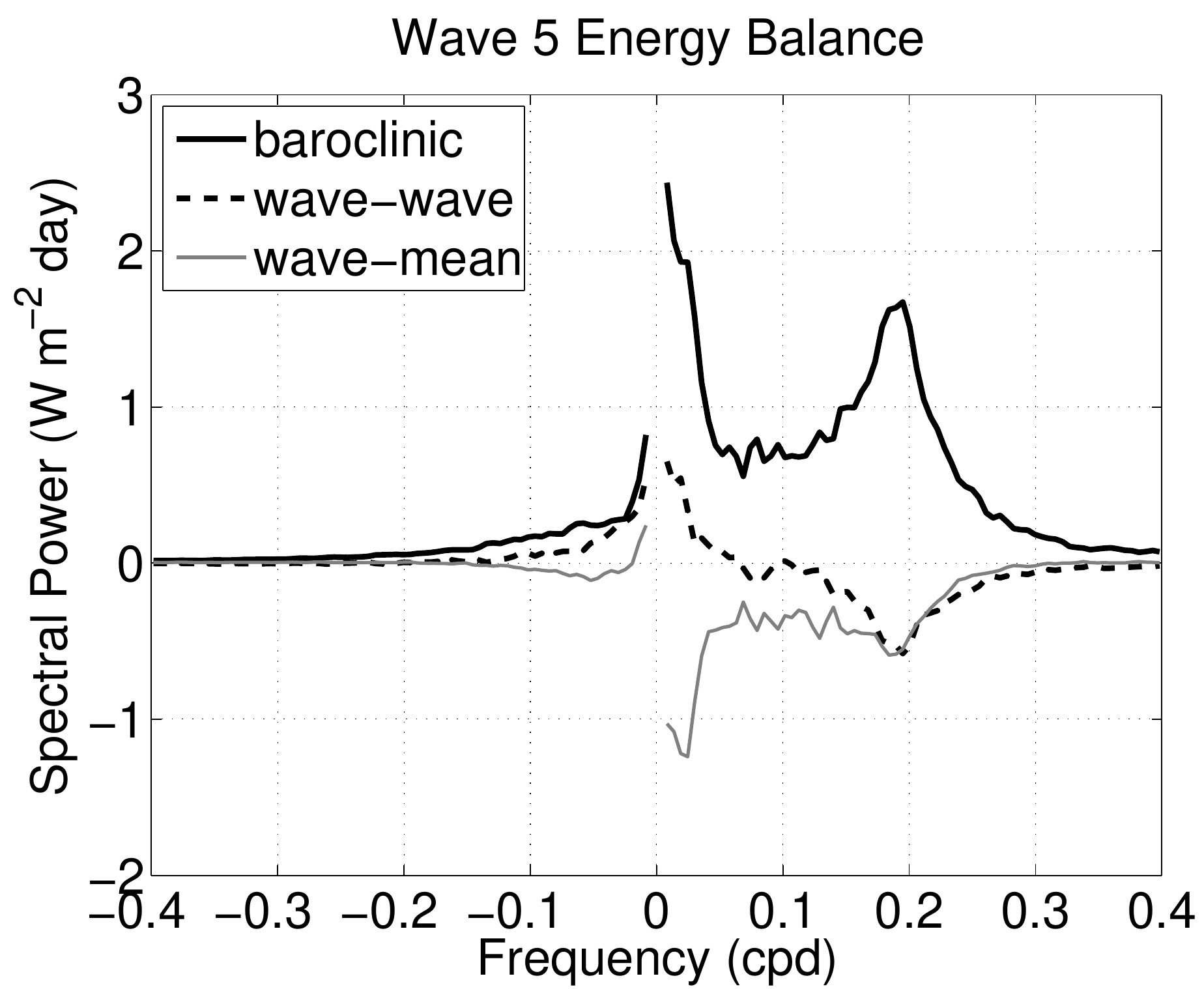}
\caption{Spectral distribution of kinetic energy sources for waves with k=5 as a function of frequency by fundamental atmospheric processes. Contributions by direct baroclinic energy conversion (heavy full line), non linear wave--wave interactions (heavy dashed line) and barotropic wave--mean flow processes (light full line) are plotted in the figure. See the text for a description of the methodology. Units are in $\mathrm{W\,m^{-2}\,day}$  \label{fig:wv5_balance}}
\end{figure}

\begin{figure}[t]
\noindent\includegraphics[width=0.5\textwidth]{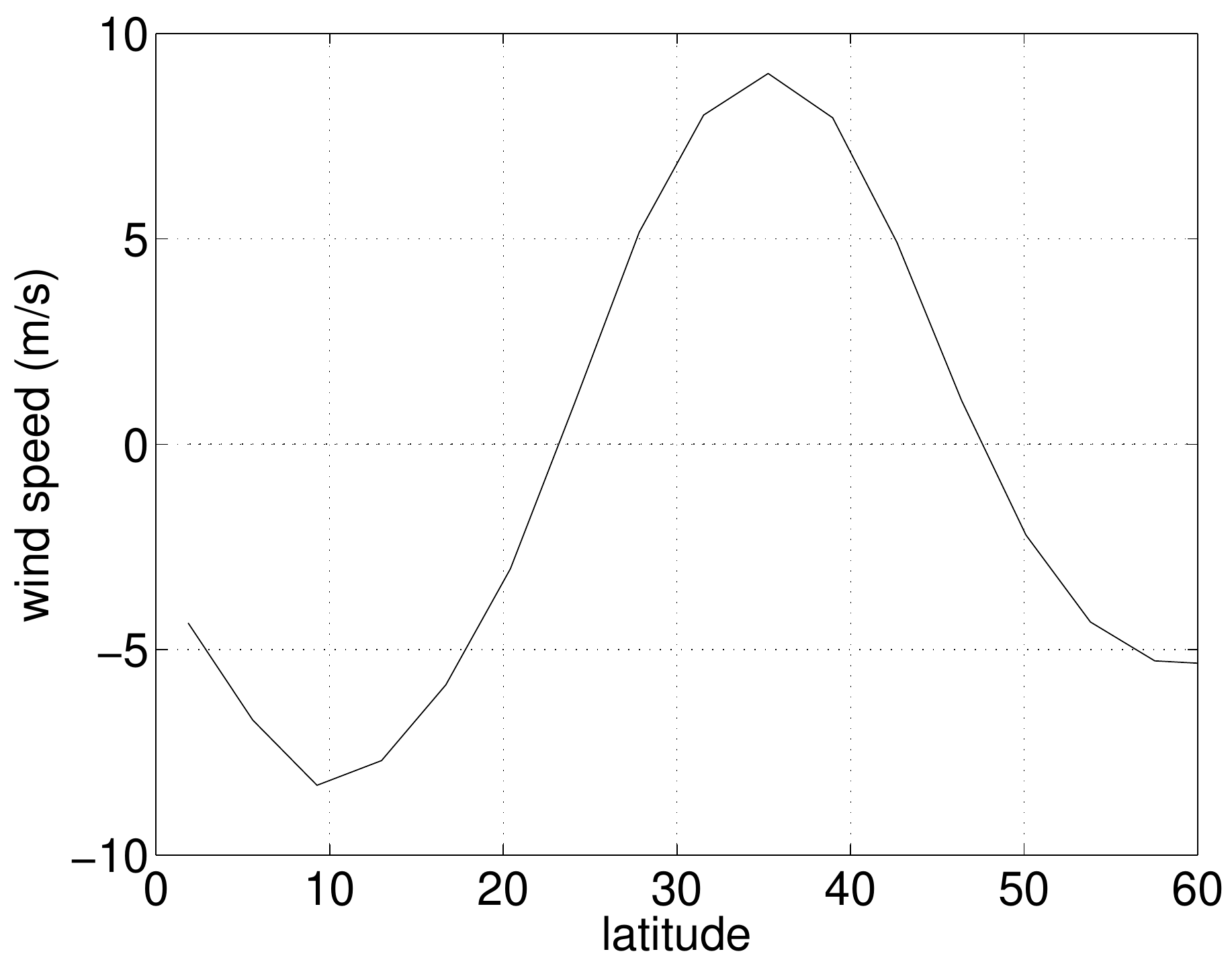}
\caption{Meridional cross section of $[\bar{u}]$ at the lowest model level. The maximum in the surface zonal wind corresponds with the latitudes where the wave five has a more barotropic vertical structure and a weaker baroclinic energy conversion.   \label{fig:ll_wind}}
\end{figure}

\begin{figure}[t]
\noindent\includegraphics[width=0.5\textwidth]{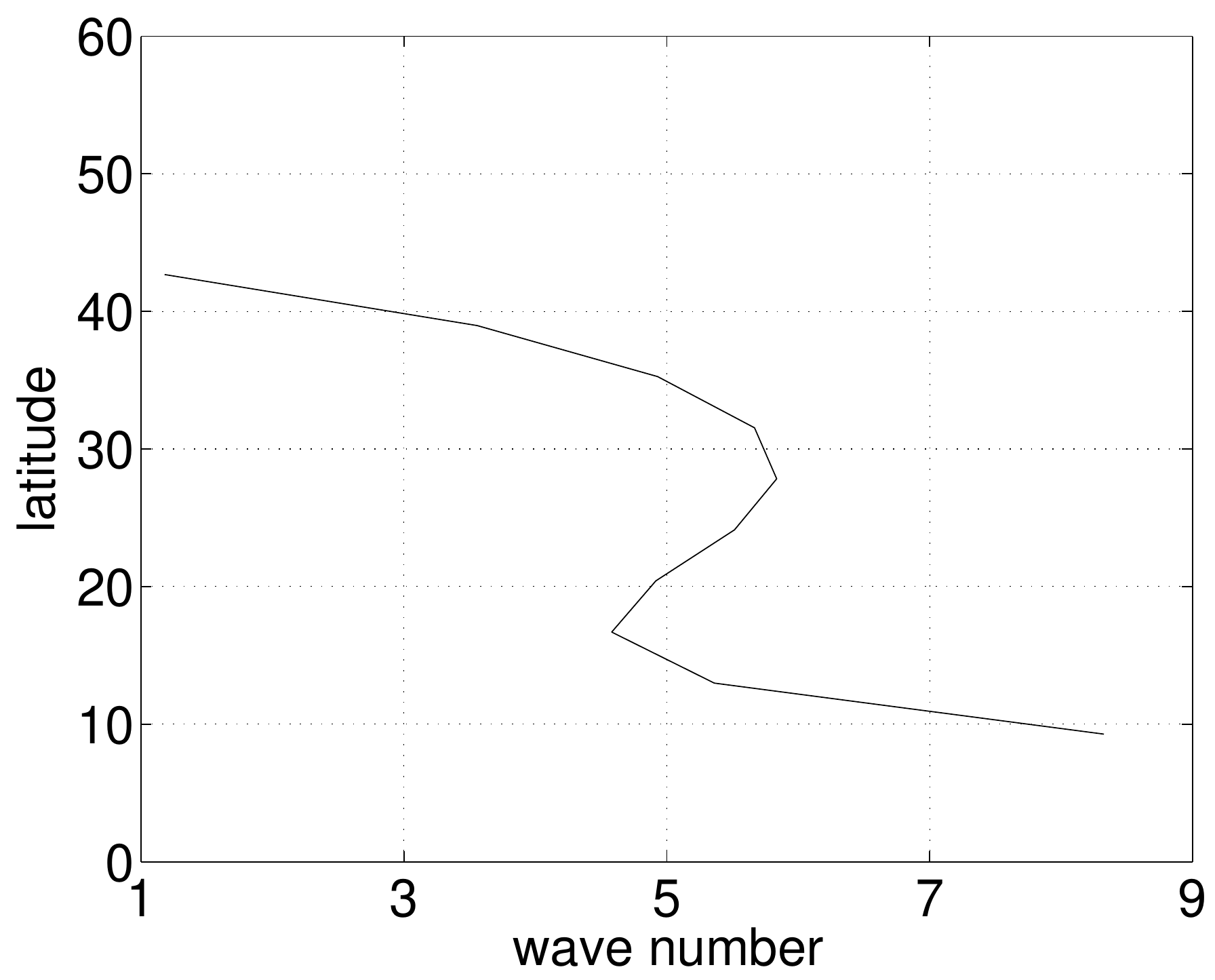}
\caption{Stationary Rossby wavenumber at 200mb as a function of the latitude. The increase observed around $30^\circ$ is due to the meridional curvature of the zonal flow at the jet stream,  which creates an upper tropospheric wave guide for a stationary Rossby wave five. 
\label{fig:ks_rossby_dt27}}
\end{figure}

\begin{figure}[t]
\noindent\includegraphics[width=0.9\textwidth]{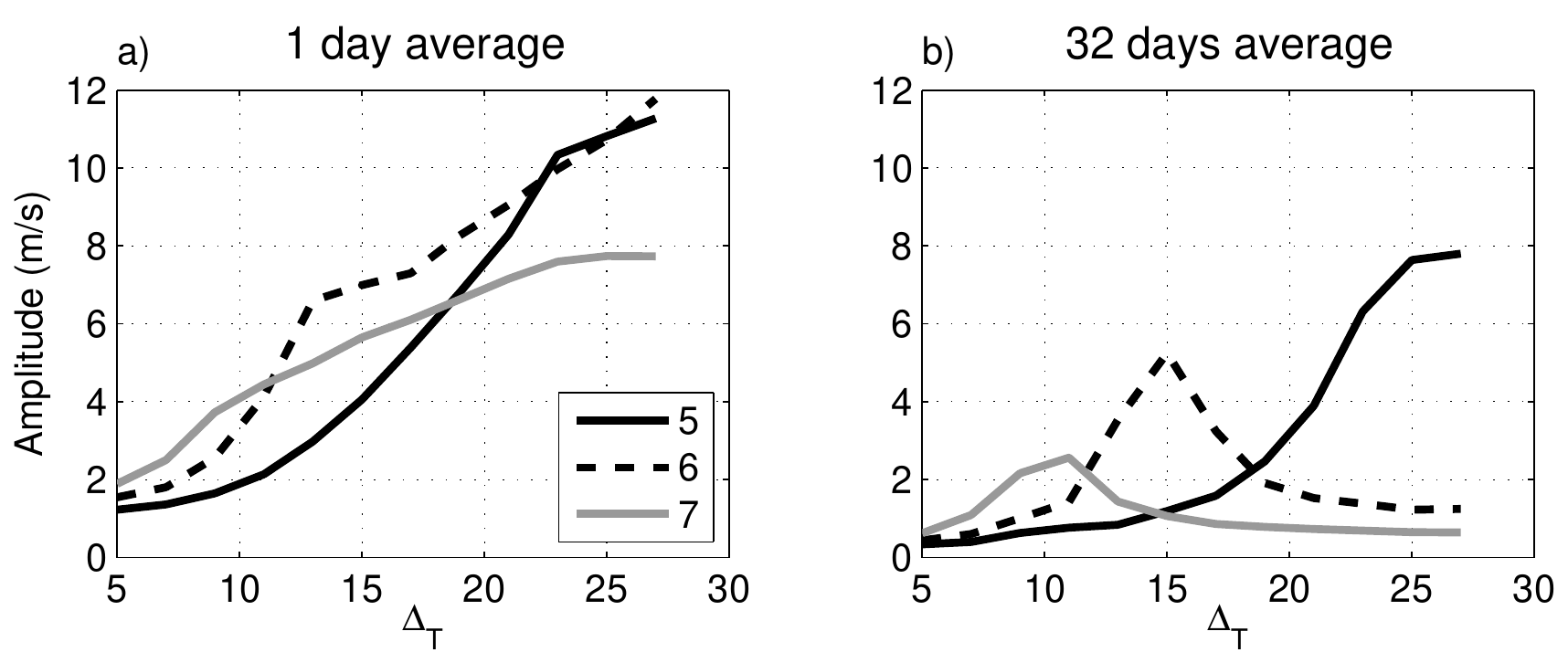}
\caption{Mean amplitude of the zonal waves $k=5$, $k=6$ and $k=7$ in $\hat{V}$ as a function of the equator to pole SST difference ($\Delta_T$). $\hat{V}$ has been filtered,  as described in Fig.~\ref{fig:amplitude},  by averaging over time windows of length (a) 1 day and (b) 32 days, so that a measure of the total and low frequency waves amplitude is respectively selected. \label{fig:amplitude_dt}}
\end{figure}

\begin{figure}[t]
\noindent\includegraphics[width=0.6\textwidth]{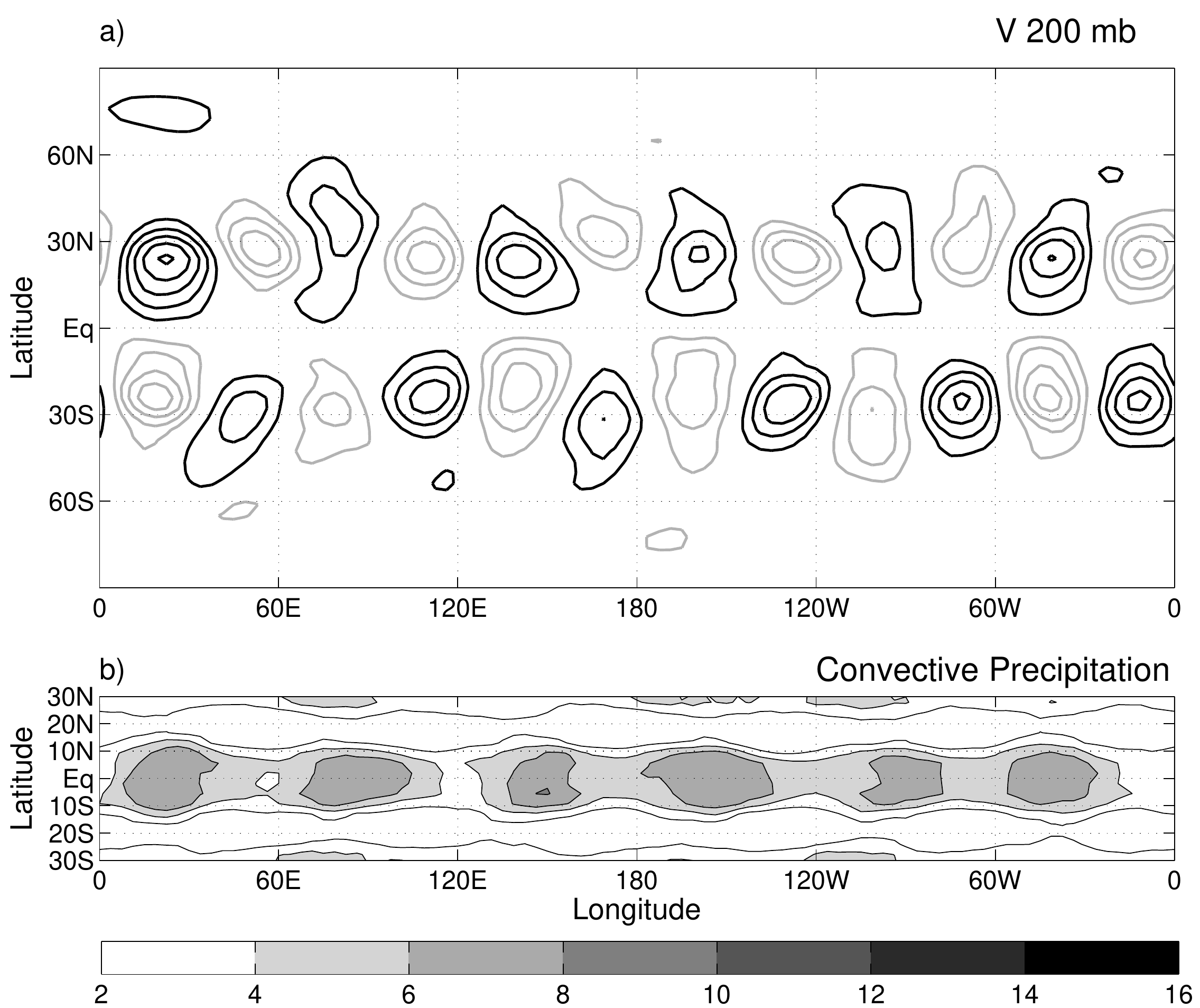}
\caption{Same fields shown in Fig.~\ref{fig:example_dt27} for a 6 month average taken from the aquaplanet simulation with SST parameters $T_e=27, \Delta_T=15$. As the baroclinicity of the system decreases, the wave number six becomes quasi--stationary in place of the previously observed wave five.  \label{fig:example_dt15}}
\end{figure}

\begin{figure}[t]
\noindent\includegraphics[width=0.9\textwidth]{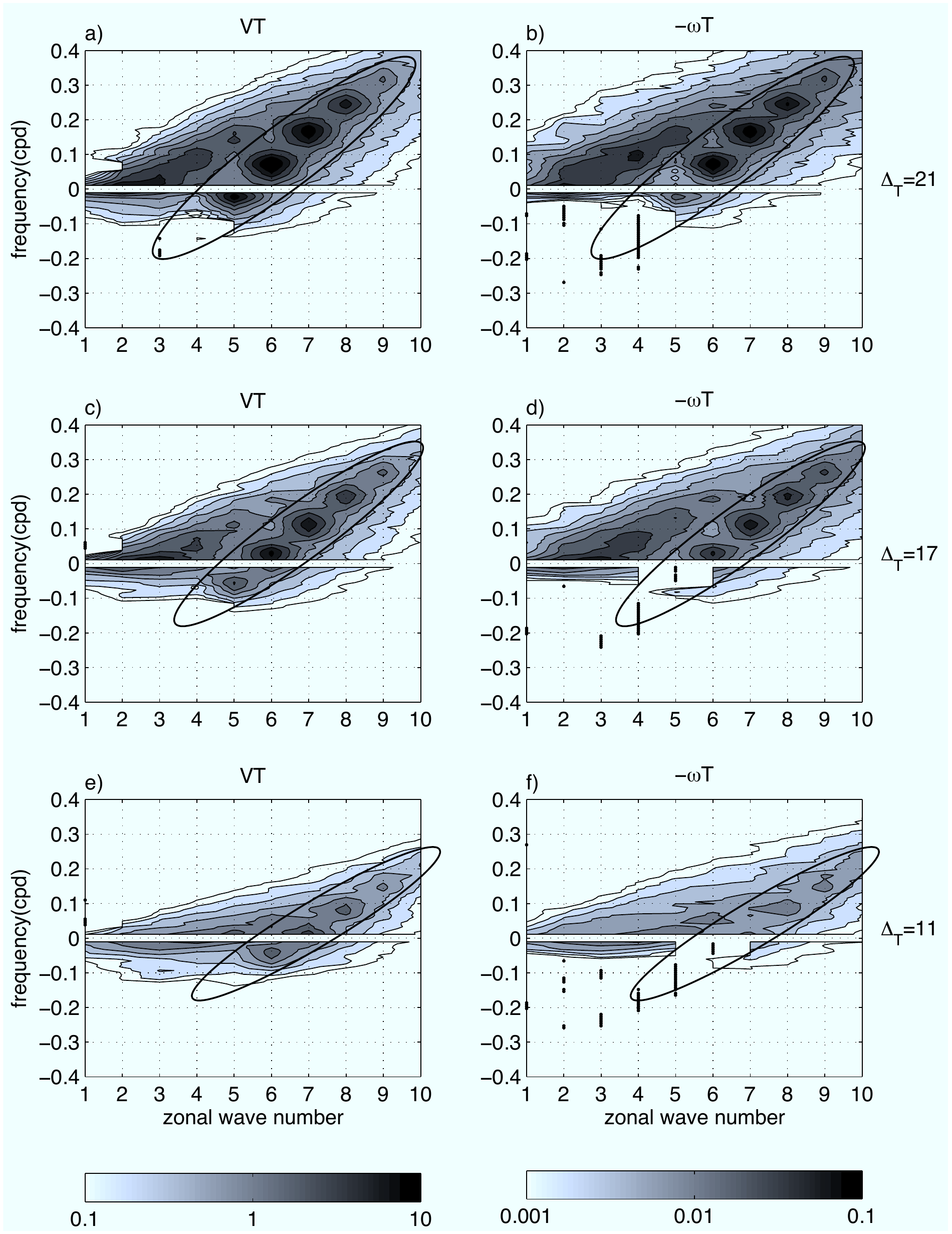}
\caption{As in Fig.~\ref{fig:cospectra_dt27} but for simulations differing in the equator to pole SST difference: (a) $\Delta_T=21^\circ$C, (b) $\Delta_T=17^\circ$C and (c) $\Delta_T=11^\circ$C.  The equatorial SST is kept fixed at $27^\circ$C. As $\Delta_T$ decreases the wave five becomes stable, and the waves six (b) and seven (c) become the new marginally stable quasi--stationary baroclinic wave. \label{fig:cospectra} }
\end{figure}
%

%%%%%%%%%%%%%%%%%%%%%%%%%%%%%%%%%%%%%%%%%%%%%%%%%%%%%%%%%%%%%%%%%%%%%
% ACKNOWLEDGMENTS
%%%%%%%%%%%%%%%%%%%%%%%%%%%%%%%%%%%%%%%%%%%%%%%%%%%%%%%%%%%%%%%%%%%%%
\begin{acknowledgment}
 We'd like to acknowledge D.~Williamson for introducing us to the problem of the wave five in aquaplanets, and M.~Blackburn, B.~Hoskins, J.~Methven and A.~Speranza for useful comments on the set up of the experiments and the interpretation of the results. We also thank an anonymous reviewer whose suggestions helped to improve the paper. VL acknowledges the financial support provided by the FP7-ERC project NAMASTE (Ref. No.: 257106).
\end{acknowledgment}

%%%%%%%%%%%%%%%%%%%%%%%%%%%%%%%%%%%%%%%%%%%%%%%%%%%%%%%%%%%%%%%%%%%%%
% APPENDICES
%%%%%%%%%%%%%%%%%%%%%%%%%%%%%%%%%%%%%%%%%%%%%%%%%%%%%%%%%%%%%%%%%%%%%
\ifthenelse{\boolean{dc}}
{}
{\clearpage}

\begin{appendix}
% Use \begin{appendix} and not \begin{appendix}[A] for only one appendix.
\section*{\begin{center} Hayashi spectra \end{center}}
\cite{Hayashi:1971} proposed a generalised method to calculate the spectrum of the eastward and westward propagating waves of a longitude--time field $u(\lambda,t)$.  The method is a generalisation of \cite{Deland:1964} quadrature spectral analysis and it is based on a zonal Fourier transform followed by a quadrature spectral analysis between the temporal series of the sine and the cosine coefficients of the zonal expansions. The Hayashi power spectrum is defined as:
\begin{equation*}
H^{\pm\nu}_{k}(u)=(P^{\nu}(C_k,C_k)+P^{\nu}(S_{k},S_{k})\pm2 Q^{\nu}(C_k,S_k))/4,\\
\end{equation*}
where $H^{\pm\nu}_{k}$ is the spectral power of the eastward ($+\nu$) and westward ($-\nu$)  propagating waves, and $P^\nu$  and $Q^\nu$ are respectively the 1D cospectrum and the quadrature spectrum on the time variable of the cosine and sine zonal Fourier coefficients of $u(\lambda,t)$:
\begin{equation}
u(\lambda,t)=u_{0}(t)+\sum_{k=1}^{N}C_k(t)\cos(k \lambda)+S_k(t)\sin(k \lambda)
\end{equation}
The spectral contributions to the longitude--time covariance ($[\overline{uu'}]$) between two fields $u$ and $u'$, can be computed by the 2D cospectrum ($P^{\pm\nu}_{k}$) :
\begin{align}\label{eq:2D_cospectra}
P^{\pm\nu}_{k}(u,u')=&(P^{\nu}(C_k,C'_k)+P^\nu(S_k,S'_k)\pm  \nonumber \\ 
                      &Q^\nu(C_k,S'_k)\mp Q^\nu (S_k,C'_k))/4.
\end{align}
Each spectrum is obtained as an average of 40 spectra computed on non overlapping 6 month long time windows (both hemispheres are considered), and by further averaging over 3 neighbouring frequency bins. The variance of the estimator of the power spectral density (PSD) is computed from the standard deviation of the 120 (3*40) spectral amplitudes that are averaged in the estimation of the PSD at each spectral bin. The one dimensional quadrature and co spectrum have been calculated by the FFT method.

Hayashi's formulas have been criticised by \cite{Pratt:1976} because standing waves are not resolved but are seen as a couple of eastward and westward  propagating waves of equal amplitude, thus limiting the only meaningful quantity to the difference between the power in the eastward and westward components. A variety of different approaches were thus developed to overcome this problem \citep{Pratt:1976,Hayashi:1977,Fraedrich:1978}. Nevertheless we decided to attain to the first Hayashi formulation because in aquaplanet models there is  no preferential phase and standing wave activity is of limited interest. Standing variance would just contain, depending on the formulation, noise or spectral power of quasi--stationary waves alternating an eastward to a westward propagation. 

\end{appendix}

%%%%%%%%%%%%%%%%%%%%%%%%%%%%%%%%%%%%%%%%%%%%%%%%%%%%%%%%%%%%%%%%%%%%%
% REFERENCES
%%%%%%%%%%%%%%%%%%%%%%%%%%%%%%%%%%%%%%%%%%%%%%%%%%%%%%%%%%%%%%%%%%%%%
% Create a bibliography directory and place your .bib file there.
\ifthenelse{\boolean{dc}}
{}
{\clearpage}
\bibliographystyle{ametsoc}
\bibliography{./bibliography/references}

\end{document}